\documentclass[preprint,authoryear,12pt]{elsarticle}

 \usepackage{graphicx}
\usepackage{amssymb}

\usepackage[
a4paper=true,%
breaklinks=true,%
colorlinks=true,%
pdfauthor={P. Mazzei et al.},%
pdftitle={Template for manuscripts in Advances in Space Research}%
]{hyperref}

\journal{Advances in Space Research}

\begin{document}

%%%%%%%%%%%%%%%%%%%%%%%%%%%%%%%%%%%%%%%%%%%%%%%%%%%%%%%%%%%%%%%%%%%%%%%%%%%%%
%% Frontmatter
\begin{frontmatter}

\title{Catching Spiral - S0 transition in groups. \\
Insights from SPH simulations \\ with chemo-photometric implementation}
%\tnoteref{footnote1}}
%\tnotetext[footnote1]{This template can be used for all publications in Advances in Space Research.}

% Use optional labels to link authors explicitly to addresses:
\author[1]{P. Mazzei}
\author[2]{A. Marino}
\author[1]{R. Rampazzo}
\author[2]{G. Galletta}
\author[1]{D. Bettoni}
 \address[1]{INAF-Osserv. Astronomico di Padova, Vic. dell'Osservatorio 5, 35122 Padova, Italy}
 \address[2]{Dip. di Fis. ed Astron. ``Galileo Galilei'', Vic. dell'Osservatorio 3, 35122, Padova, Italy}

\begin{abstract}

We are investigating the co-evolution of  galaxies within groups combining 
multi-wavelength photometric and  2D kinematical observations. Here
we focus on S0s showing star formation in 
ring/arm-like structures. We use  smooth particle hydrodynamical simulations (SPH) 
with chemo-photometric implementation which provide dynamical and morphological information 
together with the spectral energy distribution (SED) at each evolutionary stage. 
As test cases, we simulate the evolution of two such S0s: NGC 1533 and NGC 3626.

The merging of two halos with mass ratio 2:1, initially just composed of DM and gas, 
well match their  observed SEDs, their surface brightness profiles and their overall 
kinematics. The residual star formation today ``rejuvenating'' 
the ring/arm like structures in these S0s is then a mere consequence of  a major 
merger, i.e. this is a phase during the merger episode.
The peculiar kinematical features, e.g. gas-stars counter rotation in NGC 3626,
depends on the halos initial impact parameters. 
Furthermore, our simulations  allow to follow, in a fully consistent way,  the transition of these S0s
through the green valley in the NUV$-$r  vs. M$_r$ colour magnitude diagram,
 which they cross in about 3-5 Gyrs, before  reaching their current position in the 
 red sequence.  We conclude that a {\it viable} mechanism driving the evolution of
S0s in groups is of gravitational origin.
\end{abstract}

\begin{keyword}
 Galaxies: evolution \sep Galaxies: interactions  \sep Galaxies: individual: NGC 3626, NGC 1533
\sep Method: numerical

\end{keyword}

\end{frontmatter}

\parindent=0.5 cm

%%%%%%%%%%%%%%%%%%%%%%%%%%%%%%%%%%%%%%%%%%%%%%%%%%%%%%%%%%%%%%%%%%%%%%%%%%%%%
%% Main text
\section{Introduction}

The investigation of the co-evolution of  galaxies and groups in the nearby Universe is of
great cosmological interest since more than half of galaxies reside in 
groups.  Since
the velocity dispersion of galaxies in groups is low, i.e. comparable to the stellar velocity dispersion, mergers of
galaxies are  highly favoured with respect to clusters. 
Nearby groups provide a close-up view of phenomena 
driving the morphological and star formation (SF) evolution of  galaxy members before they 
fall into clusters \citep[e.g][]{Boselli06,Wilman09,Just10}.
  
Among open questions,  the old controversy about the {\it nature} vs. {\it transformation} 
origin of S0 galaxies rises above the others. 
Since Es and  S0s are the typical inhabitants of nearby clusters,
at expennse of Spirals (Sps), a debate about the  transformation of Sps into S0s in such environments arose. 

The processes  possibly driving the Sp$\rightarrow$S0 transformation are, indeed, still uncertain and 
debated \citep[e.g.][]{Bekki09}. They include feedback from AGN \citep{Schawinski09}
which may shut down SF, environmental effects which reduce the HI reservoir of Sps, \citep{Boselli06, Hughes09,Cortese09}
and the interplay between cold gas flows and shock heating  \citep{Dekel06}.
\citet{Wetzeletal12} find that galaxies in groups and clusters experience no significant environmental effects until they cross
within the virial radius of a more massive host halo.
\citet{Boselli06} claim that the interaction with the intergalactic medium is not the origin of the cluster S0s, which 
likely form by Sps through gravitational interaction.
Studying 116 X-ray selected galaxy groups at redshift 0.2-1,  \citet{Getal13} conclude that strangulation and disk fading  are insufficient to explain the observed
morphological dependence on environment, and that galaxy mergers or close tidal encounters must
play a role in building up the population of quenched galaxies with bulges seen in dense environments
at low redshift.

The comparison of clusters at 
$z \approx$ 0.1 - 0.2 with some at intermediate distance, $z \approx$ 0.4 - 0.5, 
shows that a sort of  morphological conversion in the galaxy population from Sps$\rightarrow$S0s 
took place about 1 - 4 Gyr ago \citep{Fasano2000}. 
This transformation anti-correlates  with the local density, i.e., in low
concentration clusters the transformation happened only about 1-2 Gyr ago.

The effects of the possible galaxy morphological transformation manifest themselves also
 in the  nearly ubiquitous bimodal distribution of galaxies in 
the  colour-magnitude diagram (CMD) \citep[e.g.][]{Strateva01, Lewis02, Baldry04}. 
The red disk  galaxy population  should be the result of transformations of the blue, 
late-type galaxies via mechanisms of gas depletion and consequently of SF quenching. 
 The galaxy transition from star forming to quiescent,
is highlighted by the presence of an intermediate zone,
the so--called {\it green valley} (GV), on CMDs  \citep{Martin07, Wyderetal07, Fang12}.  
Investigating galaxies in the GV should shed light on the mechanisms governing 
the {\it on-off} state of the SF. S0s in groups, in particular, are  key elements in
identifying both the mechanisms and the environmental influence  on their possible transformation.

Although S0s are  early-type galaxies (ETGs), so they are widely considered 
 evolved ``red \& dead'' galaxies, {\it GALEX (Galaxy Evolution 
Explorer)}  far-UV (FUV) and/or H$\alpha$ images 
revealed  signatures of on-going star formation in their disks
in the form of outer blue ring/arm-like structures in some of them. \citet[][and references therein]{Salim12} 
show a wide collection of these kind of galaxies.   Signatures of ongoing ($\approx$ 9$^{+4}_{-3}$\%) 
or recent ($\approx$47$^{+8}_{-7}$\%) SF  are also found in the nuclear regions  of nearby ETGs \citep[e.g.][]{Rampazzo13}.

\begin{figure*}
  \begin{center}
{\includegraphics[width=9.5cm]{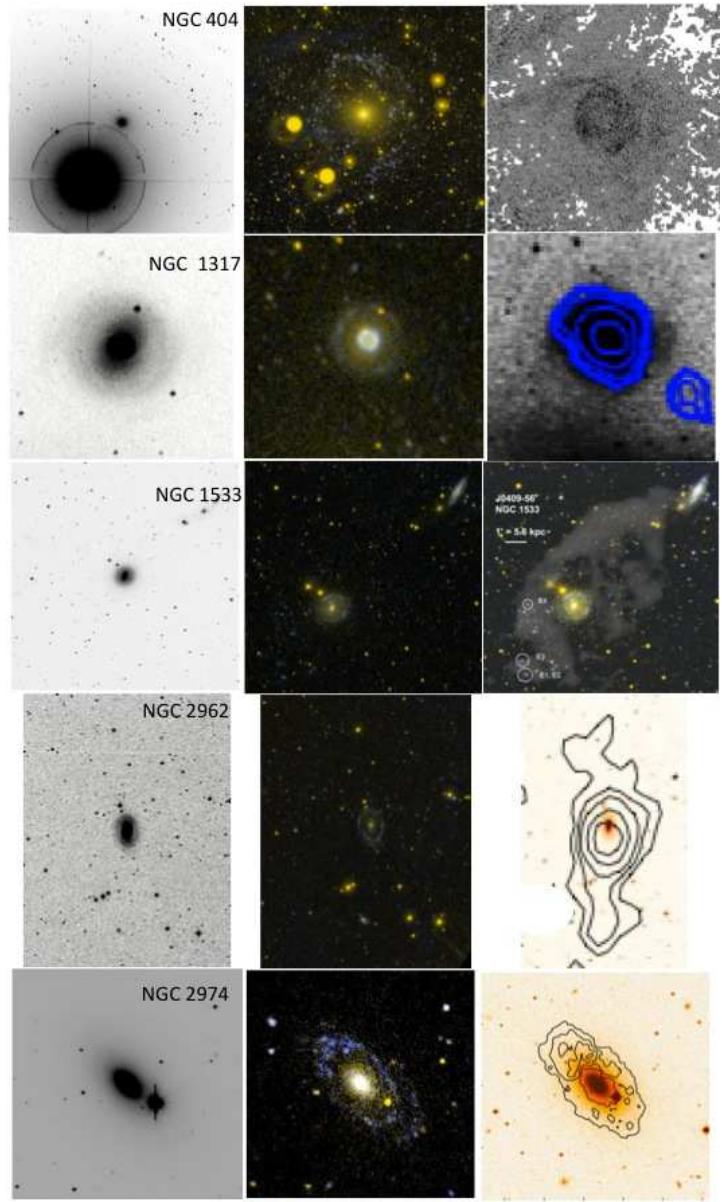}}
        \end{center}
      \caption{Composite view of five S0s with outer ring/arm-like structures detected in the 
      FUV. From left to right: optical   (left),  UV two colours FUV (blue) and NUV (yellow) composite images 
      (mid), and  HI distribution or contours (right), are shown on the same scale.
      NGC 404 images are adapted from \citet{Thilker10},  NGC 1533, NGC 2962, and
      NGC 2974 from \citet{Ant11_II}, NGC 1317 from \citet{Gil07}. HI images: NGC 1533  from \citet{Werketal10},
      NGC  1317 from  \citet{Horellou01}, NGC 2962 from \citet{Grossi09}, and NGC 2974 from \citet{Weijmans08}.
}
       \label{HI4}
   \end{figure*}

The presence of a disk and the occurrence of SF in ETGs imply that cold gas has played an
important role in their evolution.
In Figure~\ref{HI4} we show the optical (left panels) and UV
colour composite images (middle panel) of examples of nearby S0s with outer
ring/arm-like structures detected in the FUV band: NGC~404 \citep{Thilker10}, 
NGC~1533, NGC~2962,  NGC~2974 \citep{Ant11_II}, and NGC~1317 \citep{Gil07}.
All the S0s in Figure~\ref{HI4} are located in low density environments. These galaxies  also
 have cold gas, distributed either over the main body 
 of the galaxy or forming large-scale structures around the galaxy, 
 as shown by  HI contours over plotted in the right panels of Figure~\ref{HI4}. 

%-----------------------------------  Figure 1 --------------------------------
% ---------------------------- end Figure 1 --------------------------------   
%-----------------------------------  Figure 2 --------------------------------
\begin{figure*}
  \begin{center}
     {\includegraphics[width=8cm]{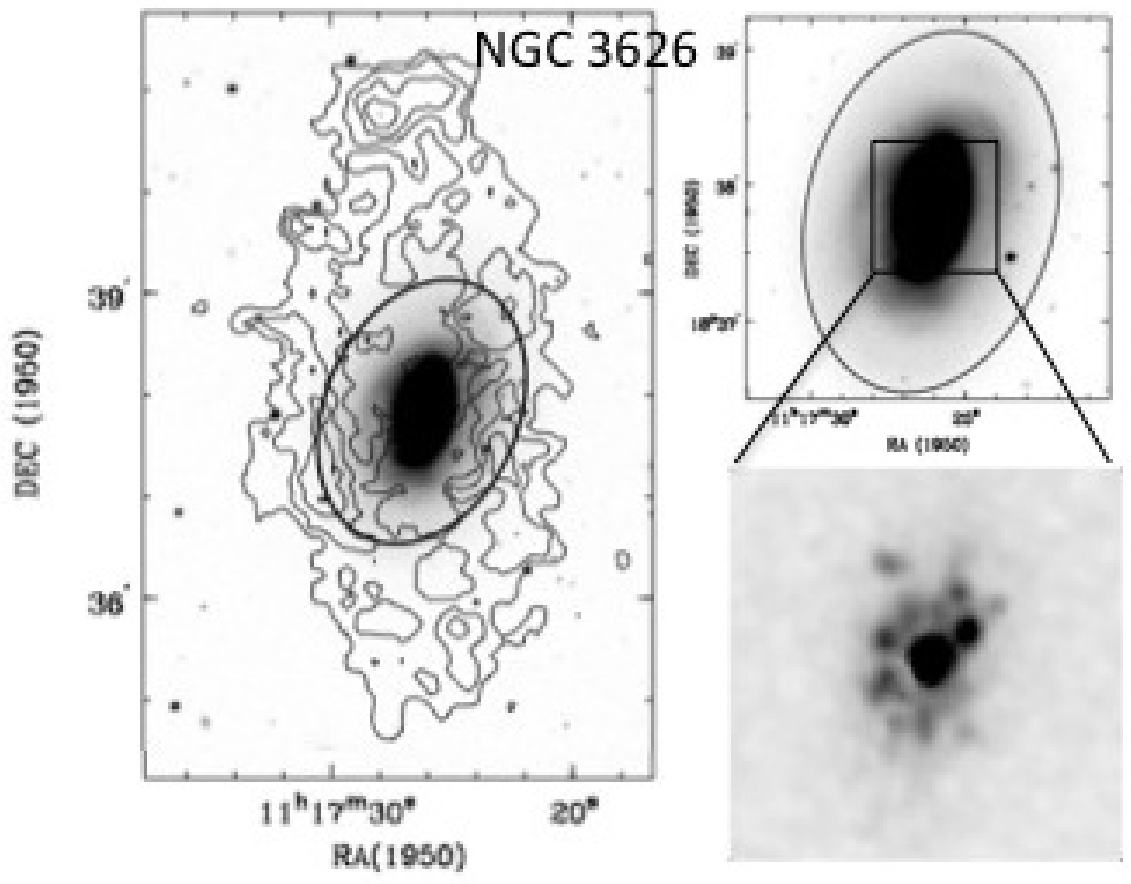}} 
        \end{center}
      \caption{ {\sl Left:} Optical (R-band) image of NGC 3626 with HI contours superposed.  {\sl Right:} Optical (top) and
      H$\alpha$ (bottom) images adapted from \citet{Haynesetal00}.
The ellipse has the dimension D$_{25} \times$ d$_{25}$ and is oriented along the optical major axis of the galaxy; d$_{25}$  is 1'.51 ({\tt NED}).}
       \label{HI4_II}
   \end{figure*}
% ---------------------------- end Figure 2 --------------------------------   

The paper explores  the evolution of these S0s, using  
SPH simulations accounting for chemo-photometric evolution, i.e providing
the SED at each evolutionary time (snapshot). 
 The SED takes into account chemical evolution, stellar  emission,
internal extinction and re-emission by dust in a self-consistent way, as already
described in MC03, \citet[and references
therein]{marilena}, and \citet{marilena2}. The SED extends over four orders of magnitude in
wavelength, i.e., from 0.1 to 1000 $\mu$m.  Each simulation
self-consistently provides morphological, dynamic and chemo-photometric
evolution.

Simulations of groups, focusing on galaxy morphological transformation have been 
approached with different techniques (e.g. \citet{Kawata08, Bekki11,Villalobos12}).
\citet{Kawata08} used a cosmological 
chemo-dynamical simulation to study how the group environment affects the SF properties
of a disk galaxy.    
 \citet{Bekki11}, starting from already formed galaxies, show that
Sps in group environments can be strongly influenced by repetitive, slow encounters
so that galaxies with thin discs can be transformed into thick disks, and gas poor S0s.
\citet{Villalobos12} perform 17 experiments in order to study the general evolution of disc galaxies,
composed of a stellar disc embedded in a DM halo, within a group environment. This is modelled as an N-body DM halo
following a \citet{NFW97} density profile including a spherically symmetric stellar component at its centre.

None of the quoted simulations explore the merger scenario combining the chemo-dynamical code with  evolutionary  
population synthesis (EPS) models
providing the galaxy SED at each evolutionary time (snapshot hereafter).
This upgrade in galaxy simulations is the novelty of this paper. 

 As a case study we focus on simulations of  two S0s with ring/arm like structures,
detected in the far UV and/or narrow band optical images, immersed in extended HI emission, 
namely NGC~1533 (Fig. \ref{HI4}) and NGC~3626 (Fig. \ref{HI4_II}). 
Both galaxies are members of groups \citep{Ant12,Ant11}. 
NGC~3626, is located within the virial radius of USGC~U376  (\citet{Rametal02}, Mazzei et al. 2013, in prep.)
and NGC~1533 is a member of the Dorado group.  They have extended SEDs, from FUV to far-IR (FIR), 
that we will use, together with their B-band absolute magnitude, to constrain our simulations. 
The detailed surface photometry and the 2D kinematical properties of the gas and 
stars of NGC~3626, also available from the literature, will be  a further  constraint.

The SPH simulations best-fitting their global properties are selected 
over a large grid of encounters from halos initially composed of DM and gas. In addition, 
our simulations highlight  the evolutionary path through the GV in the CMD, of such galaxies. 

The structure of the paper is as follows.  Section 2 describes  the observed  properties of
NGC~3626 and NGC~1533.   Section 3 provides the details of our grid 
of SPH simulations. In Section 4 we compare the global measured galaxy properties 
with the results of our simulations. Section 5 discusses the galaxy evolution 
provided by the simulation  focusing on the  path of the near-UV(NUV)/optical 
colours in the ({\it NUV-r}) vs. M$_r$ CMD. In Section 6 we draw our conclusions.

%------------------------    Table 1 -----------------------------------------------
\begin{table*}
\caption{Structural, photometric and kinematics properties of  NGC~1533 and NGC~3626}
\begin{tabular}{cccccccccc}
\hline
Name & T  &Dist   &$D_{25}$ &$r_e$ & M$_B$  & $\Delta V_{gas}$ \\
  &  & [Mpc]  &[kpc] &[kpc] & [mag] & [km~s$^{-1}$]\\
\hline
\hline
NGC~1533&   SB$0-$ & 26.1 $\pm$1.9 &10.3& 1.6 &-19.59 $\pm$0.38 & 132.1 $\pm$ 6.3 \\%ok
NGC~3626&    (R)SA(rs)$0+$ & 11.0 $\pm$0.8 &22.3& 3.3 &-20.11 $\pm$0.42  & 191.1 $\pm$ 5.3 \\%ok

\hline
\end{tabular}
\\
\label{tab:obs_e}
\end{table*}
%------------------- end Table 1 --------------------------------------------------

\section{Photometric and kinematic  properties of NGC 1533 and NGC 3626}

The relevant properties of NGC~1533 and NGC 3626  are  summarized in Table
\ref{tab:obs_e}: columns 2 and 3  give the morphological type and the 
cosmological distance  from {\tt NED\footnote{NASA/IPAC Extragalactic Database http://ned.ipac.caltech.edu}};
column 4 the apparent diameter corrected for galactic extinction and inclination,
 $D_{25}$, from {\tt HYPERLEDA}\footnote{http://leda.univ-lyon.fr1}, \citep{Patetal03};  column 5 the effective radius, r$_e$ from
the RC3 catalogue \citep{DV91},   column 6 the  absolute B-band magnitude  
derived from the mean distance modulus computed from redshift-independent 
distance measurements ({\tt NED}), and column 7 the maximum rotation velocity corrected for inclination  from {\tt HYPERLEDA}.
$D_{25}$ and $r_e$ are reported in kpc using  cosmology
corrected scales from {\tt NED}  with cosmological parameters  
H$_0$=70\,km/s/Mpc, $\Omega_{matter}=$0.27, $\Omega_{vacuum}$=0.73.

{\bf NGC~1533} is a SB$0-$ in the {\tt RC3} catalogue \citep{DV91}, 
a SB$02(2)/$SBa in the  {\tt RSA} catalogue \citep{Sandage87}, and a $(L)SB(rs)0^0$
in the  {\tt NED} database. This latter classification  
implies a  large lens outside the bar and an inner ring \citep{Laurietal06}.
\citet{Ant11_II} found a prominent outer ring  in the UV images extending 
from 39$^"$ to 80$"$, and  consisting of young ($\leq$200 Myr old) stellar
populations.  More details on  {\tt GALEX} observations are given in  \citet{Ant11}.
\citet{Werketal10} found a large-scale distribution of HI around the galaxy, 
as shown in Fig. \ref{HI4}.  \citet{RW03} reported HI
arcs around the galaxy, spanning from 2$'$ to 11$'$.7  outside the FUV ring, with a mass of
9.3 $\times$ 10$^8$ M$_{\odot}$ (their Table A1). They concluded that the HI is most 
likely the remnant of a tidally disrupted galaxy. This HI mass estimate increases by 
at least 6 times  adopting the distance in \citet{dG07}. 
From  optical spectroscopy in the central region of this  galaxy, \citet{Ann07}  derived an age of 11.9$\pm$6.9 Gyr. 
\citet{dG07} describes  NGC~1533 as a galaxy which is completing a transition  from late to early 
type: it is red, but not completely dead. \\

{\bf  NGC 3626} is  a lenticular galaxy with an outer (quite faint) ring (RLAT+..) 
in the RC3 catalogue \citep{DV91}, a normal Sa  in the RSA  catalogue \citep{Sandage87} 
and a normal S0  galaxy in the {\tt NED} classification (Table \ref{tab:obs_e}). 
It has inner and outer rings starting from an arm-like structure

Photometric \citep{Sietal10, AS07, Laurietal05} and kinematic observations 
\citep{Cetal95, G98, Haynesetal00} show that its  structure  and kinematics are 
quite complex. Figure \ref{HI4_II} illustrates its wide scale and inner properties.
A wide dust-lane west of the nucleus is clearly visible  within 1' 
\citep[their Fig. 5]{Haynesetal00}.
Ionized gas and CO disks counter-rotate with respect to stars  \citep{Cetal95,G98}. 
The stellar vs. ionized gas velocity profiles are 
followed out to 45", i.e., 0.5~$R_{25}\sim$6\,kpc, and 30" respectively, on both sides 
of the galaxy  \citep{Haynesetal00}.
HI VLA data  \citep{Haynesetal00} show that the cold gas extends about 3' each side of the optical nucleus.
The HI kinematics confirms the  large-scale  decoupling of the stars and gas. 
Evidence of an additional HI gas component co-rotating with the stars
is suggested by \citep{Haynesetal00}, as hinted also in the $^{12}$CO map by \citet{G98}.
$^{12}$CO emission is concentrated in a compact nuclear disk of average 
radius r=12". From $R$=20" to $R$=100" $^{12}$CO is not detected, 
and the neutral gas content is largely dominated by HI. The HI in the inner regions 
avoids the nucleus, where H$\alpha$ emission dominates,  but follows closely 
the H$\alpha$ outside \citep{Haynesetal00}.

 \citet{Laurietal05}, from a Fourier analysis of the
 K$_s$-band image of the galaxy, suggest the presence of a 40" long bar, 
 and of an inner ($<$5") elliptical structure, likely a secondary bar or a disk structure
 with a twisted (10$^{\circ}$) position angle with respect to the main disk.
 \citet{Sietal10} find indications of two
stellar disks. The inner dsik extends from 10" to 45" with  a scale
length of 19".8$\pm$0".7  ($i$ SDSS band), and the 
outer one has a scale length of 21''$\pm$2". The bulge 
is very compact with an effective radius of 2".52$\pm$0".05.
 \citet{Sietal10} derived an age younger than 2\, Gyr in the  central part of NGC~3626, suggesting the occurrence of a recent accretion episode.

The calibration of \citet{Bell03} and \citet{Hop03} allows us to derive  the star formation rate (SFR)  from the radio flux (1.4 GHz).
Adopting a luminosity distance of 26.1\,Mpc  and the  measure of \citet{cond02} at 1.4\,GHz, the SFR of stars more massive than 
5\,M$_\odot$ is  $\simeq$1.7 M\,$_\odot$/yr. 
This becomes  0.5 M\,$_\odot$/yr  using  \citet{Crametal} calibration.

\section{The grid of SPH chemo-photometric simulations}

Our SPH simulations of galaxy formation and evolution start from
the same initial conditions as described in \citet[][MC03 hereafter]{paola5}
 and \citet{CM99} i.e. a collapsing  triaxial systems initially composed of
dark matter (DM) and gas in different proportions and different total
masses. In more detail, each system is built up with a spin parameter,
$\lambda$, given by $|{\bf {J}}||E|^{0.5}/(GM^{0.5})$, where E is the
total energy, J the total angular momentum and G the gravitational
constant;  $\lambda$ is equal to 0.06 and aligned with the shorter
principal axis of the DM halo. The triaxiality ratio of the DM halo,
$\tau=(a^2-b^2)/(a^2-c^2)$, where $a>b>c$, is 0.84 \citep{war92}.

All the simulations  include self--gravity of gas, stars and DM,
radiative cooling, hydrodynamical pressure, shock heating, artificial
viscosity, star formation (SF), feedback from evolved stars and type
II SNe, and  chemical enrichment as in MC03 (and references therein).
{\it They provide the synthetic SED at each evolutionary step, i.e., at each
snapshot, accounting for evolutionary population synthesis models}.
 The SED takes into account chemical evolution, stellar  emission,
internal extinction and re-emission by dust in a self-consistent way, as
described in previous works (\citet{marilena} and \citet[and references
therein]{marilena2}). This extends over four orders of magnitude in
wavelength, i.e., from 0.1 to 1000 $\mu$m. Each simulation
self-consistently provides morphological, dynamic and chemo-photometric
evolution. The Initial Mass Function (IMF) is of Salpeter type with
upper mass limit 100\,$M_\odot$ and lower mass limit 0.01$\,M_\odot$
\citep{Salp55} as in  \citet{CM99} and MC03. All the
model parameters adopted here  had been tuned in previous papers  where
the integrated properties of simulated galaxies, i.e., colours, absolute
magnitudes, mass to luminosity ratios provided by different choices of
model parameters after 15 Gyr, had been successfully compared with those
of local galaxies (see also \citet{paola1}; \citet{paolaa}). In
particular, from this IMF choice a slightly higher SFR  arises compared with the other possibilities examined;
this allows for the lowest feedback strength, and for the expected
rotational support when disk galaxies are formed \citep{CM99, paola5}. Moreover, as pointed
out by \citet{K12}, its  slope is almost the same as the Universal Mass
Function which links  the IMF of galaxies and stars to those of brown dwarfs,
planets and small bodies (meteoroids, asteroids) \citep{BH07}.

With respect to MC03, the particle resolution is enhanced here to
4-8$\times$10$^4$ instead of  1-2$\times$10$^4$, so there are
2-4$\times$10$^4$ particles of gas and 2-4$\times$10$^4$ of DM at the
beginning in each new simulation. Moreover the time separation between the snapshots has been halved; it is now 37\,Myr.
The gravitational softenings are 1, 0.5, and 0.05\,kpc respectively for DM, gas,
and star particles.

A new large set of  galaxy encounters involving systems with 1:1 and
different mass ratios have been performed  with  the  same initial
conditions as described in MC03. By seeking to exploit a wide range of
orbital parameters, we carried out different simulations for each couple
of interacting systems varying the orbital initial conditions in order
to have, for the ideal Keplerian orbit of two points  of given masses,
the first peri-center separation,  $p$, equal to the initial length of
the major axis of the more massive triaxial halo down to 1/10 of the same axis. For
each peri-center separation we changed the eccentricity in order to have
hyperbolic orbits of different energy. The spins of the systems are
generally parallel each other and perpendicular to the orbital plane, so
we studied direct encounters. Some cases with misaligned  spins
have been also analysed in order to investigate the effects of the system
initial rotation on the results. For a given set of encounters
with the same orbital parameters  we also examined the role of
increasing initial gas fractions.

\section{Comparing data  with  models}

From the grid of physically motivated SPH simulations, we isolate
those  simultaneously  best-fitting the global properties, i.e., 
the total SED, the B-band absolute magnitude, and the photometric and kinematics
properties of NGC 1533 and NGC 3626,  presented in \S~2.
The selected snapshot corresponds to the best-fit age of the galaxy, given in the following.
 
We find that both the galaxies  are well matched by {\it  the same simulation
at different evolutionary times}.

The simulation consists of  a  merger of two
triaxial collapsing  systems initially composed of DM and gas with  mass ratio 2:1 and total mass 
3.0$\times10^{12}$ M$_\odot$. The starting point  corresponds to
6$\times$10$^4$ particles, 3$\times$10$^4$ of gas and 3$\times$10$^4$ of
DM  with a relative mass gas ratio of 0.1. The gas mass resolution is
10$^{7}$ M$_\odot$, that of the DM particles  $\approx$9 times larger.
The spins of the systems are equal ($\lambda$=0.06, MC03),
 perpendicular, and both aligned with the shorter of their
principal axes. The initial parameters  of the encounter are reported in 
Table \ref{tabinput}. In particular the first peri-centre separation,
$p$,  corresponds to 1/10 of the major axis of the principal system, the
orbit eccentricity is 1.3, and the anomaly corresponds to 200$^{\circ}$.
The variables $r_1$, $r_2$, $v_1$, $v_2$ in Table \ref{tabinput} are the initial 
positions and velocities with respect to the centre of mass of the system of
the two halos having masses M$_1$ and M$_2$.

In the following,  we discuss in detail the comparison of our models with
the  available data for the two galaxies discussed in  \S 2. 

\subsection{ NGC~3626}

The galaxy age is of 11.5\,Gyr. The
average stellar age within the effective radius, $r_e$ provided in Table \ref{tab:obs_e},  
is 4.3\,Gyr and increases to  5.5\,Gyr  within $R_{25}\simeq$~3 $r_e$. 
These age estimates become  younger if  stellar ages are 
weighted by the B-band luminosity: 3.4\,Gyr and 4.5\,Gyr respectively. 
In the inner regions, $R\le$1.5 ~kpc, stars are younger than 2\,Gyr, in 
good agreement with the findings of \citet{Sietal10}.

The SFR, i.e. the mass of stars younger than 0.01\,Gyr, is 
2\,M$_{\odot}$/yr, in agreement with the radio estimates given 
by \citet{Bell03} (see  \S 2). 

%-------------------- Table 2 --------------------------------------- 
\begin{table*}
\caption{Input parameters of SPH simulation of NGC~3626 and NGC~1533}
\begin{tabular}{cccccccc}
\hline
 p  & $R_1$ & $R_2$ & v$_1$ & v$_2$ & M$_1$ & M$_2$ & f$_{gas}$ \\
&  [kpc] & [kpc] & [km~s$^{-1}$] & [km~s$^{-1}$]  & [$10^{10}\,M\odot$] &[$10^{10}\,M\odot$] & \\
\hline
\hline
 101  & 273  & 546 &52 &104 & 200 & 100 & 0.1  \\
\hline
\end{tabular}
\\
\label{tabinput}
\end{table*}
%-------------------- end Table 2 ----------------------------------

%-----------------------------------  Figure 1 --------------------------------
\begin{figure*}
\begin{tabular}{cc}
{\includegraphics[width=6cm]{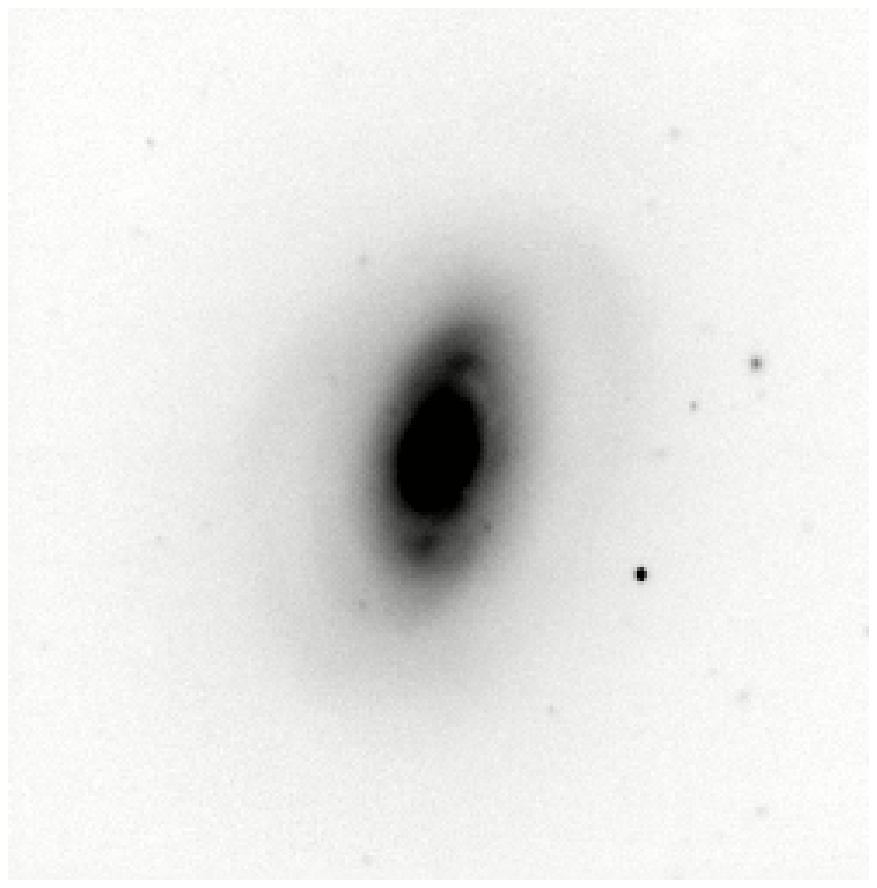}} &{\includegraphics[width=6.5cm]{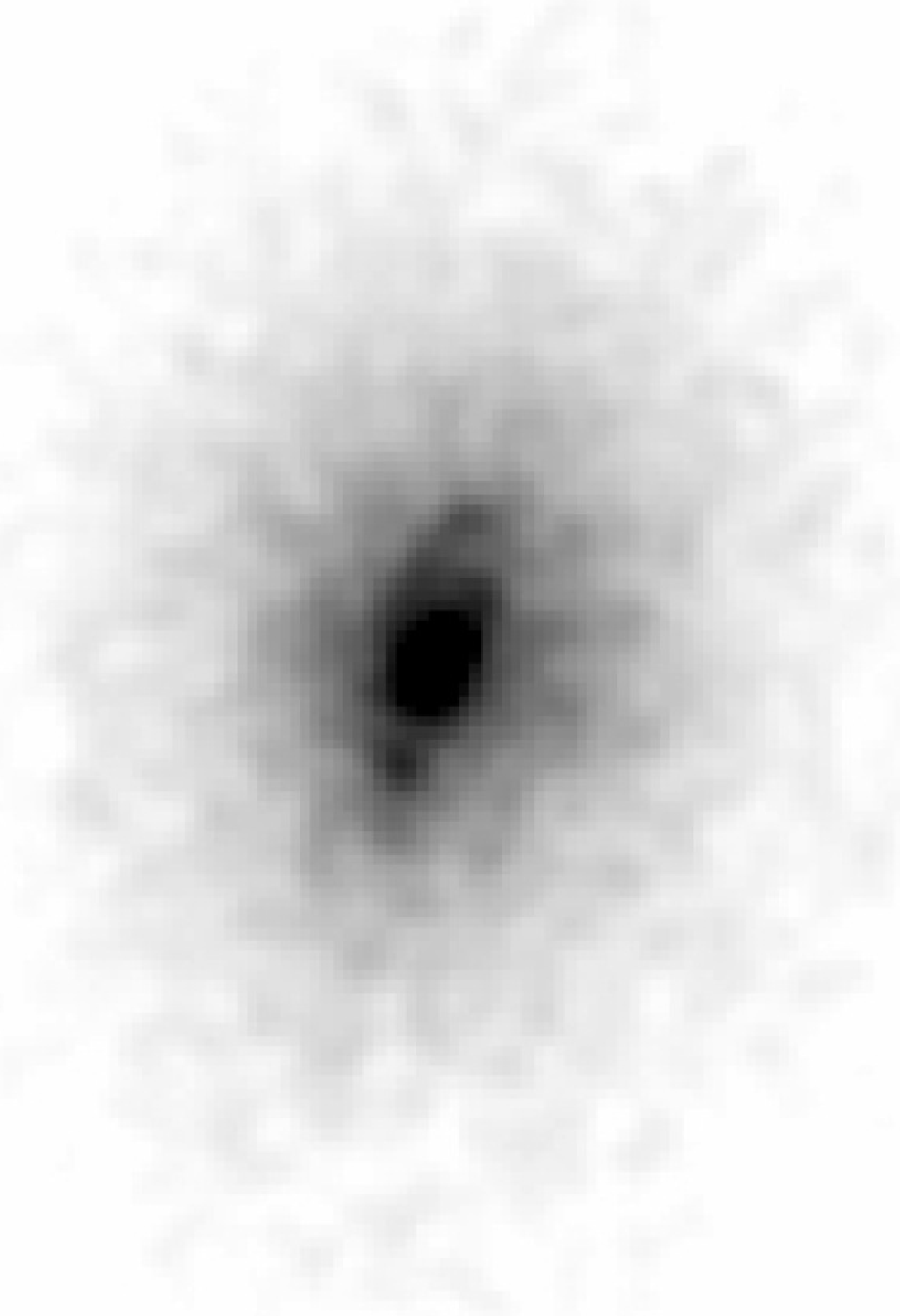}}\\
{\includegraphics[width=7cm]{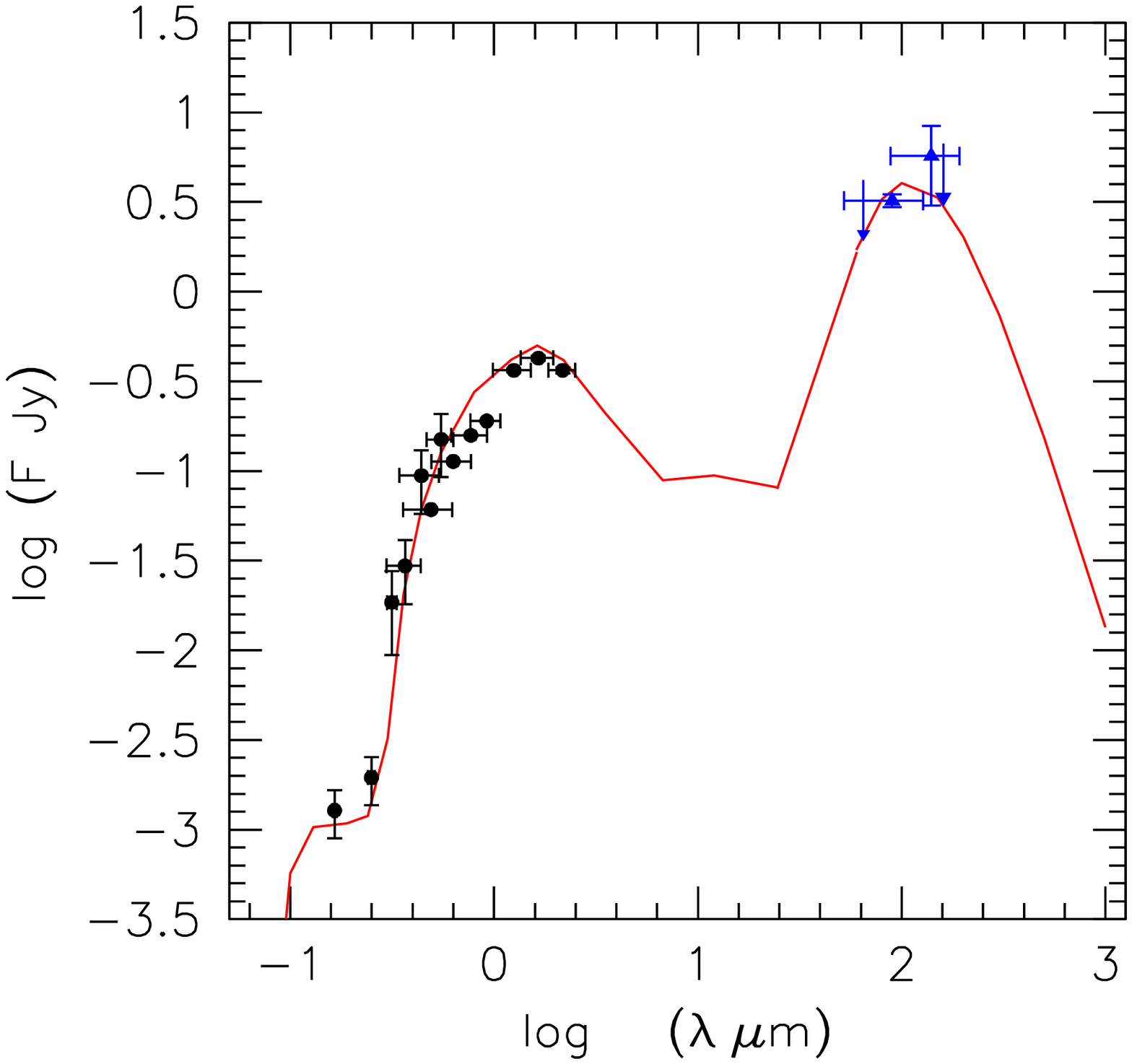}}&{\includegraphics[width=7.5cm]{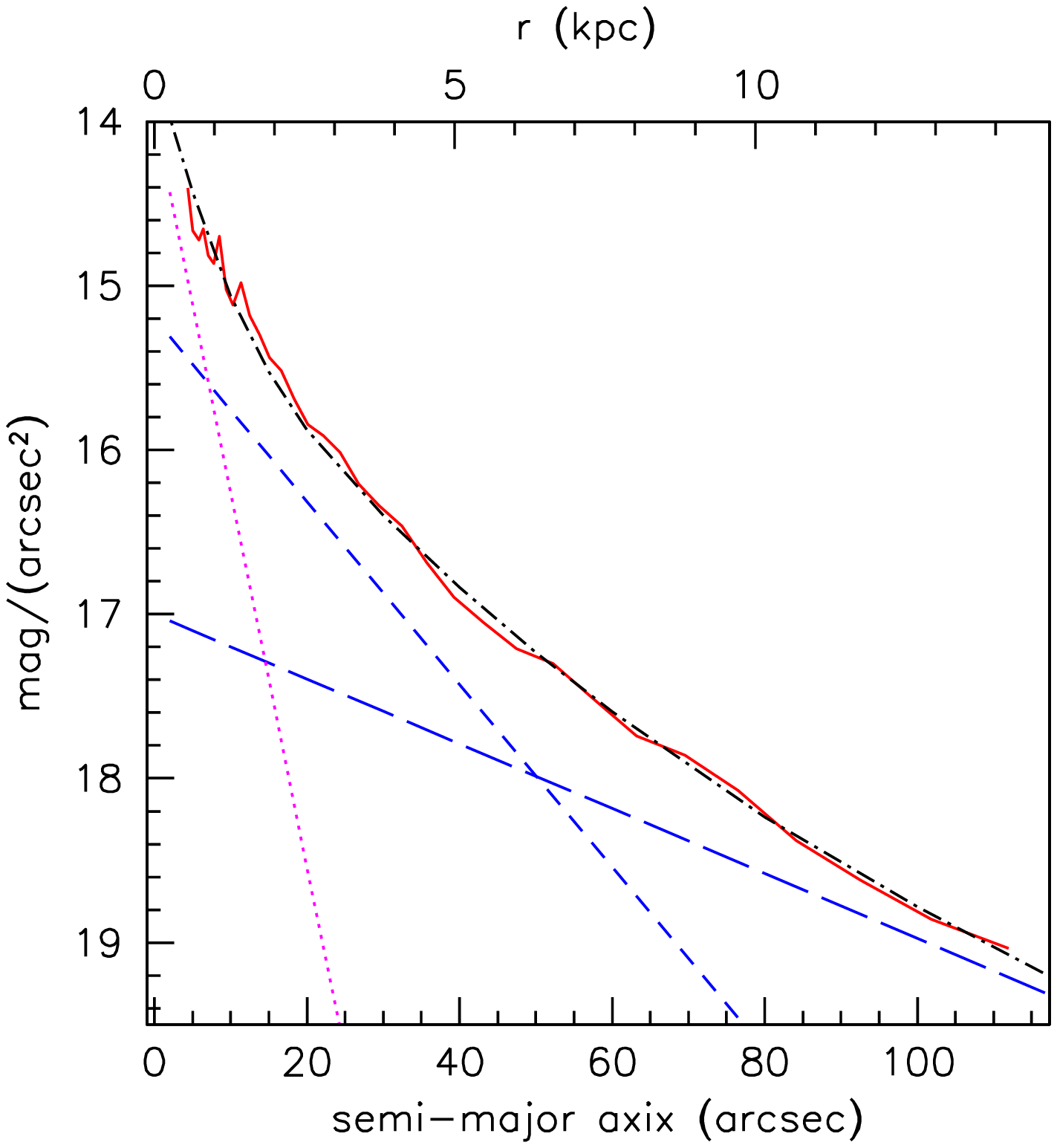}}
\end{tabular}
      \caption{
       {\sl Top:}  2.9' $ \times$2.9' J-band image of NGC~3626 (left), and
       simulated J-band map (right) on the same scale (resolution is 2"/px). 
      {\sl Bottom left:}  Red solid line  shows the prediction of our model for NGC~3626, 
filled circles are data from FUV to NIR spectral range {\tt NED}, (blue) triangles in the FIR are from the AKARI/FIS Bright Source Catalog \citep{Yamaetal09};
 error bars account for band width and 
      flux 3\,$\sigma$  uncertainties. {\sl Bottom right:} J-band brightness profile (red solid line) 
      of our simulated map and its best-fit (dot--dashed line); dotted line shows the bulge distribution using a 
      de Vaucouleurs law with R$_{e}$=8" and $\mu_J$(0)=15.8\,mag/arcsec$^{2}$; 
      short-dashed line: the inner disk with  scale length 19.5" and $\mu_J$(0)=15.2\,mag/arcsec$^{2}$;  
      long-dashed line:  the outer disk with  55" and  17\,mag/arcsec$^{2}$.
      }
       \label{fig1}
   \end{figure*}
% ---------------------------- end Figure 1 --------------------------------   

\subsubsection{Photometric properties}

Figure~\ref{fig1}  compares the  J-band image of NGC~3626  (top left panel) 
with the XZ  projection (top right panel) of the selected snapshot in the same band with the same spatial scale.
 Our simulation well reproduces the observed morphology.

The absolute  B-band  magnitude of the model, $M_B=-20.23$, agrees with
the observed value reported in Table \ref{tab:obs_e}. 

The entire SED is also well matched by our simulation at the selected snapshot (Fig. \ref{fig1}, left bottom panel). 
The SED refers to the total fluxes of the galaxy and
extends from FUV to 160\,$\mu$m, i.e. over almost three orders of 
magnitude in wavelength. UV band data are from \citet{Rifatto95} and  
FIR data,  at 65, 90, 140, and to 160 $\mu$m \footnote{65 and 160 $\mu$m are upper limits}, from AKARI/FIS  Bright Source Catalog \citep{Yamaetal09}.
The predicted  SED  in the FIR is composed of a warm and a cold dust  component
and includes PAH molecules as discussed in \citet{paola2}.  Warm dust is
located in regions of high radiation field,  i.e., in the  neighbourhood
of OB clusters, whereas cold dust is heated by the general interstellar
radiation field. The distribution of diffuse radiation is the
same as in the Milky Way \citep{paola2}, however its intensity,
I$_0$=30~I$_{local}$, is almost four times higher than in normal spirals.  Therefore,
the temperature distribution of the cold dust component peaks  at almost 25K instead of
20K as in the Milky Way. The warm dust temperature is not well constrained since no data are available in the
20-60$\mu$m spectral range. Our fit corresponds to a  warm dust 
 temperature of  $\sim$55\,K  and  to a warm-to-cold energy ratio of 0.23, as
in \citet{paola2}. Therefore, the FIR SED is accounted
for by the diffuse radiation of disk stars on dust grains since cold dust provides
about 80$\%$ of the FIR emission. This is not surprising given the small
bulge characterizing this S0 system: its
bulge-to-total light ratio is 0.25  \citep{Laurietal05}.
The fraction of bolometric luminosity of such a galaxy  in the FIR is 34\%,, similar to that expected for 
Spirals  and at least 10 times higher than  the average of S0 galaxies \citep{paola6}.

We derive the J band luminosity profile of our simulated image (Fig. \ref{fig1} right, bottom  panel)
using the {\tt ELLIPSE} package  \citep{Jedr87} in {\tt IRAF}.
To perform the best-fit of the  surface brightness profile,
 we follow the strategy adopted by \citet{Sietal10} using SDSS images (Sect. \S 2.). We derive the inner disk extending from 7" to 50" with scale length 19".5, 
 in good agreement with the cited paper, and the outer disk, beyond 50", 
 with larger scale length, 55". This is possible since our model is not 
 background limited. The central region is well matched by a de Vaucouleurs law, typical of a bulge (Fig. \ref{fig1}).
We point out that  the J central surface brightness of our 
 outer disk  is 17\,mag/arcsec$^{2}$. This becomes 20.7\,mag/arcsec$^2$ in the B-band
accounting for the intrinsic total  colour, (B-J)$\simeq$3\,mag,   and the average total B-band internal reddening, 
 $~$0.7\,mag (derived from our SED).
This  value is well within  the range assumed by the extrapolated central surface 
brightness  of spiral disks  \citep[see][]{PD83}.

All the photometric properties analysed, are well matched at the selected snapshot of our simulation.
 
 %----------------------------   Figure  2 --------------------------------------- 
\begin{figure*}

\begin{tabular}{cc}
{\includegraphics[width=7cm]{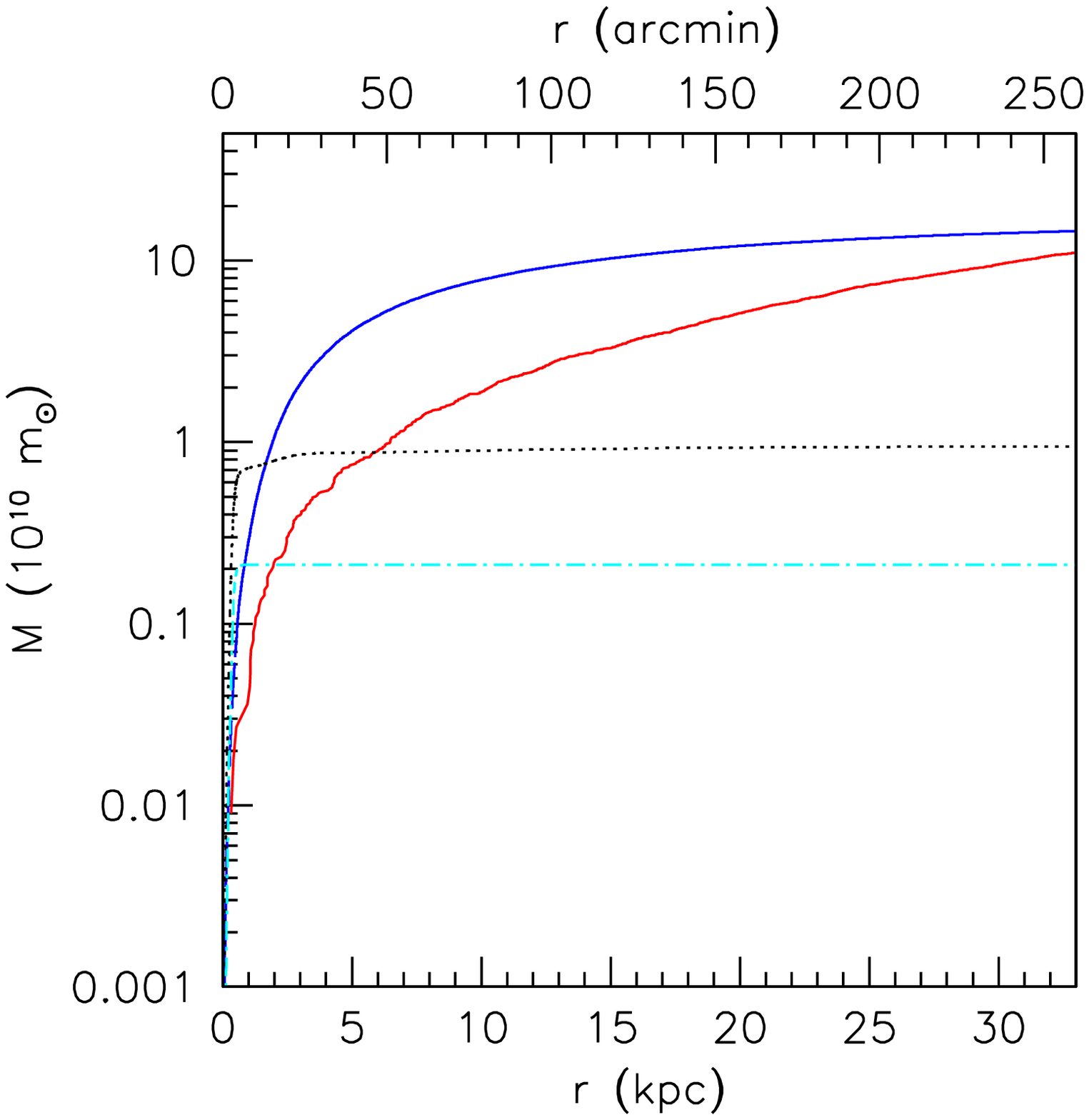}} & {\includegraphics[width=6cm]{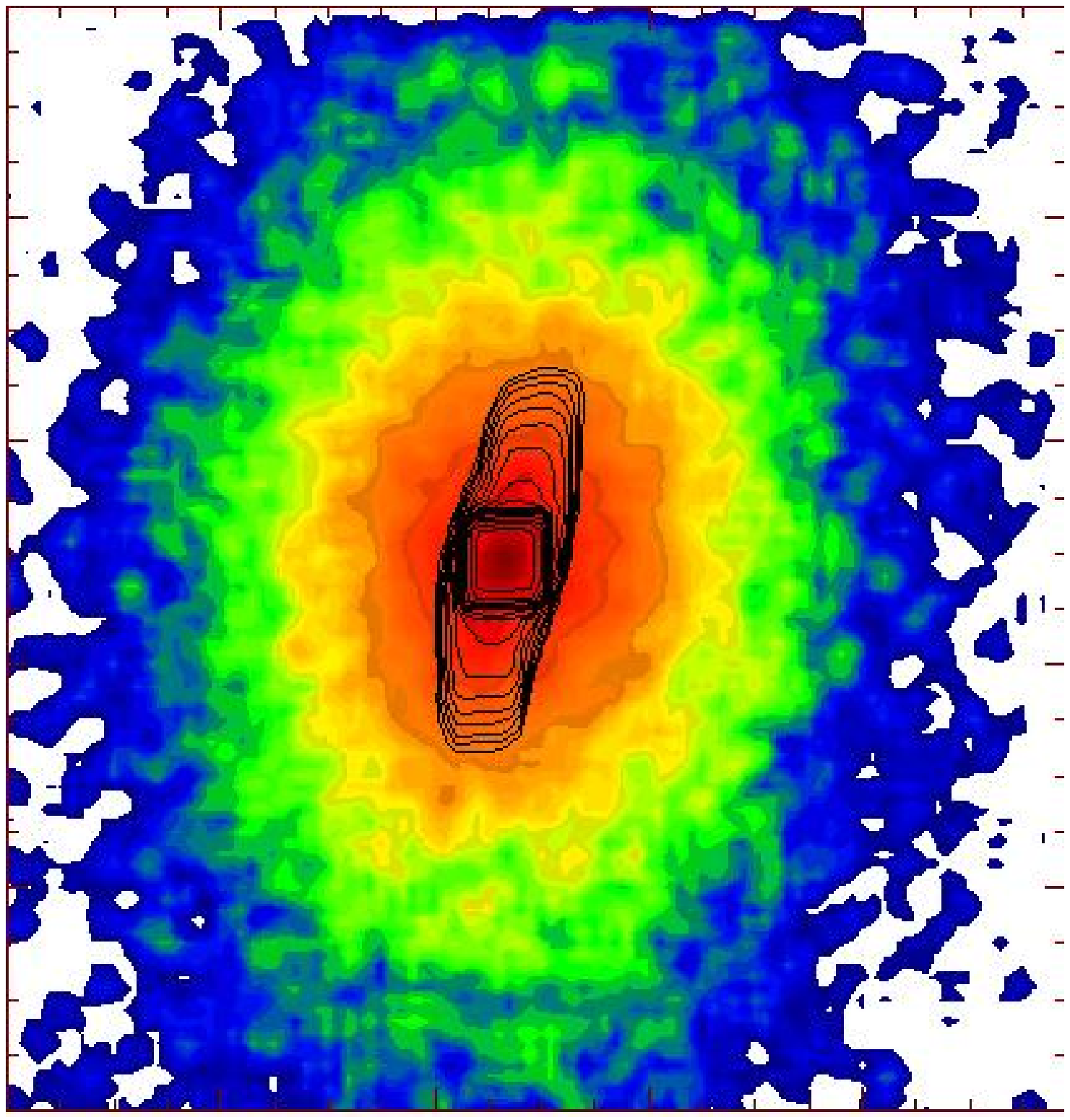}}\\ 
 & {\includegraphics[width=6cm]{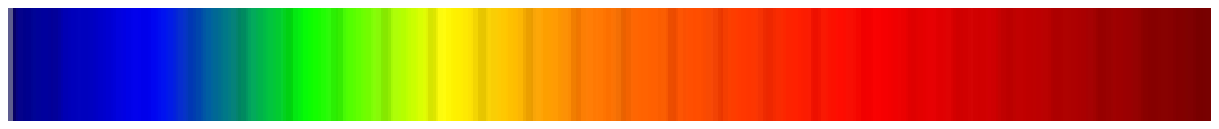}}  \\
\end{tabular}
\caption{
{\sl Left:}  Mass distribution inside a radius of 33\,kpc $\simeq$10~$r_e$ centered on the V-band galaxy
       centre: (red) solid line shows the 
       DM mass, (blue) long-short dashed line  stars, (black) dotted line  gas, 
       and (cyan) dot-dashed line  cold gas (T$<$10$^4$\,K).
    {\sl Right:} 2'.9$\times$2'.9 XZ projection in the J-band of the selected snapshot. 
    The map has been normalized to the total flux and includes  60 equally spaced levels spanning a contrast of 200; in the colour scale, 
     blue corresponds to the lower and red to the higher density contrast. Twenty equally spaced levels showing the morphology of the cold gas are over plotted on the star contours.
}
      \label{fig20}
 \end{figure*}

\begin{figure*}
\begin{center}
{\includegraphics[width=6cm]{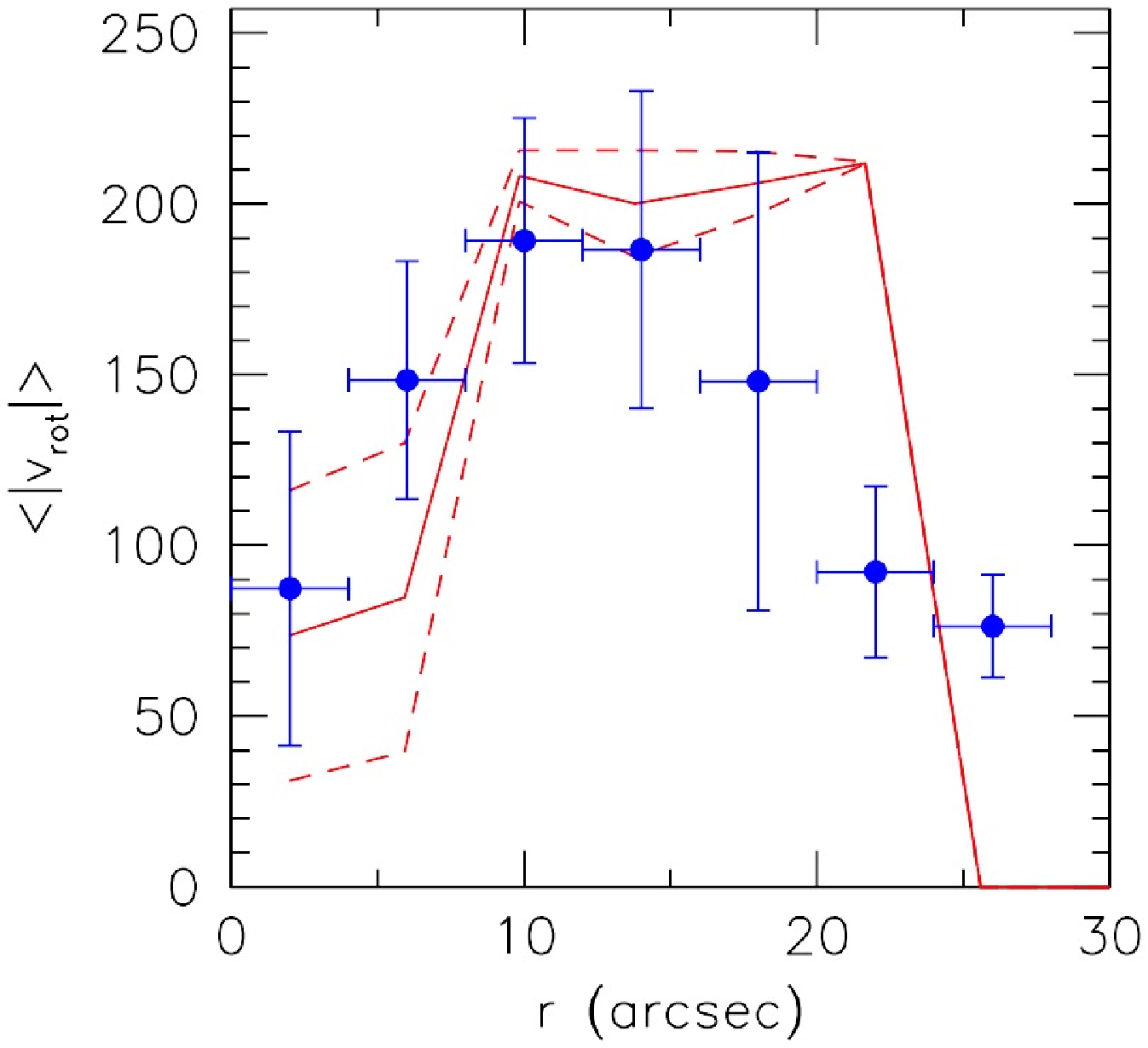}}
{\includegraphics[width=6cm]{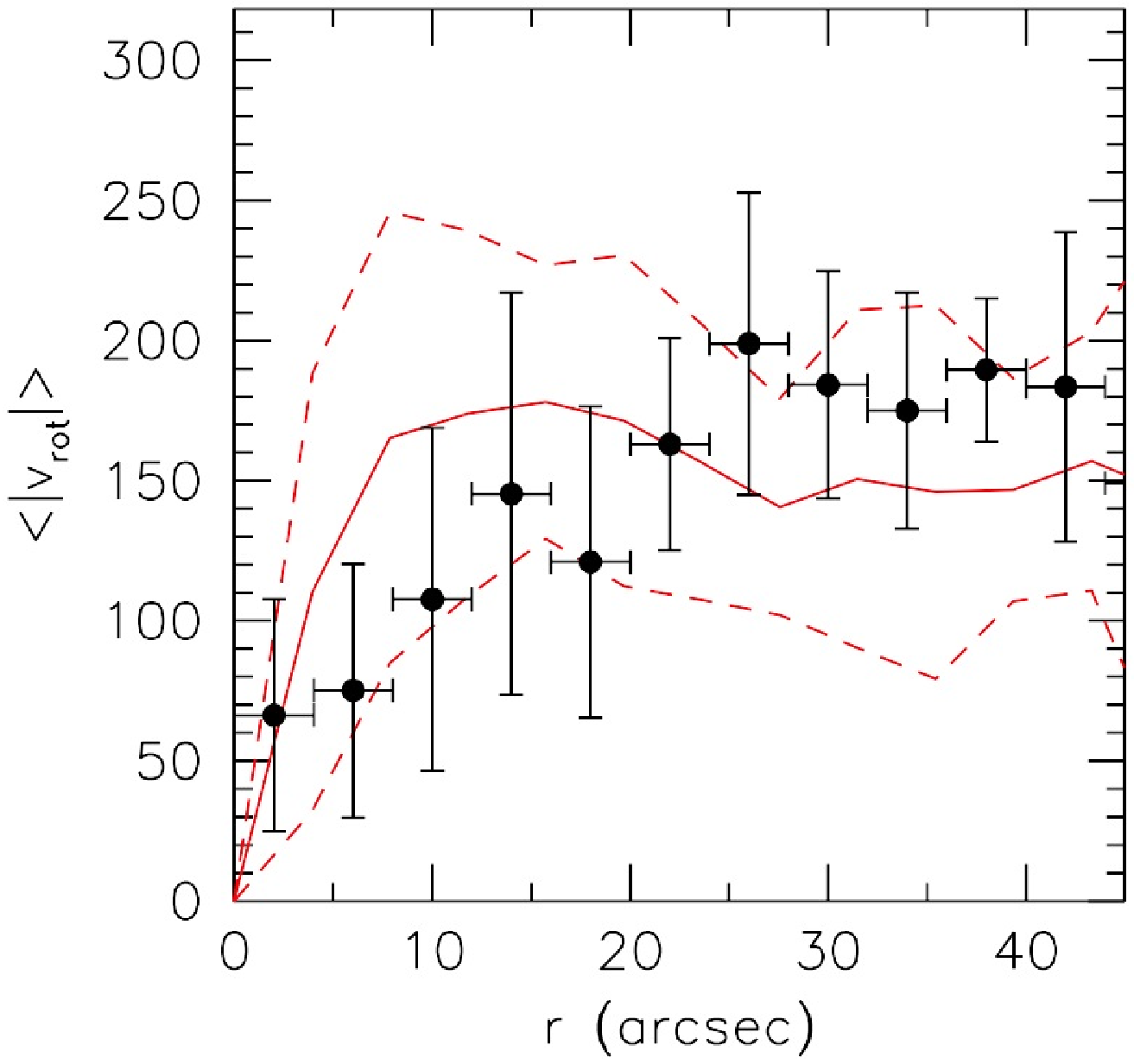}}
\end{center}
 \caption{Rotation curve of gas (left) and stars (right) compared with  our predictions.
   Filled circles, binned within 4", indicate gas measurements (left) by
 \citet{Cetal95, Haynesetal00}   and star data (right) by \citep{Haynesetal00};
 vertical bars show the velocity dispersion of observations within each bin. Dashed lines indicate the  velocity 
 dispersion  of our snapshot, showing the range of model predictions.
 }
       \label{fig21}
   \end{figure*}
% -------------------------- end figure 2 ---------------------------------------

\subsubsection{Kinematics}

Figure~\ref{fig20} (left panel) shows the distribution of the different mass components 
within  33\,kpc, i.e.,  $\simeq$10~$r_e$, at the selected snapshot. There is a large amount 
of gas inside this radius, 8.5$\times$10$^9$ M$_{\odot}$. All this gas has a temperature 
lower than 2$\times$10$^{4}$K. Since the cooling time of the gas is much shorter than the 
snapshot time range (0.037 Gyr), this amount represents the upper
limit of the mass of the coldest gas within the radius considered. In particular,  
the mass of gas with  temperature  T$\leq$10$^{4}$K,  which marks more closely the regions of SF, 
is  2$\times$10$^9$ M$_{\odot}$, in good agreement with the value of 1.1$\times$10$^9$ M$_{\odot}$  
given by \citet[their Table 8]{Haynesetal00}.

Fig. \ref{fig20} (right panel), shows the same XZ projection as in Fig. \ref{fig1} with gas contours over plotted. The disk of gas appears edge-on, rotating along the X-axis, coplanar with the stars.

The rotation curve of the cold gas (i.e. T$\leq$10$^{4}$K), is compared with the observations 
of \citet{Cetal95} and \citet[P.A. 157 degrees]{Haynesetal00} in Fig. \ref{fig21} (left panel). 
Both the data and the snapshot velocities are
binned to a spatial resolution of 4". 
We emphasize that increasing the gas temperature limit to 2$\times$10$^4$\,K, the rotation curve rises
to 25\,kpc, in agreement with the large  disk,  extending well  beyond  the optical 
radius of the galaxy (i.e., 2.6$\times$D$_{25}$), detected by \citet{Haynesetal00}.
Figure~\ref{fig21} (right panel) compares  the stellar velocity curve of our simulation, at the selected snapshot, with 
observations \citep{Cetal95, Haynesetal00}. 

Both the rotation curves  well reproduce the observed ones within the errors.

\subsection{NGC~1533}

The same simulation which best fits the global properties of NGC~3626,  also matches those
of NGC 1533 at a snapshot 2.15\,Gyr older. This corresponds to a  galaxy age of 13.7\,Gyr. 
The average stellar age within the effective radius, $r_e$ (Table \ref{tab:obs_e}), is 
6.5\,Gyr and within $R_{25}(\simeq$~3 $r_e$), is 7\,Gyr. These estimates become younger if the stellar age
is weighted by  the B-band luminosity: 3.7\,Gyr and 6\,Gyr respectively. 
The  B-band absolute magnitude, $M_B=-19.89$\,mag, agrees with the observed value 
reported in Table 1. 

The SFR, i.e. the mass of stars younger than 0.01\,Gyr, is 0.14\,M$_{\odot}$/yr, 
and the total SFR, i.e., the total mass of stars born within  the  snapshot time-step, 
is 0.26\,M$_{\odot}$/yr.

\subsubsection{Photometric properties} 
\begin{figure*}
  \begin{tabular}{cc}
{\includegraphics[width=4.5cm]{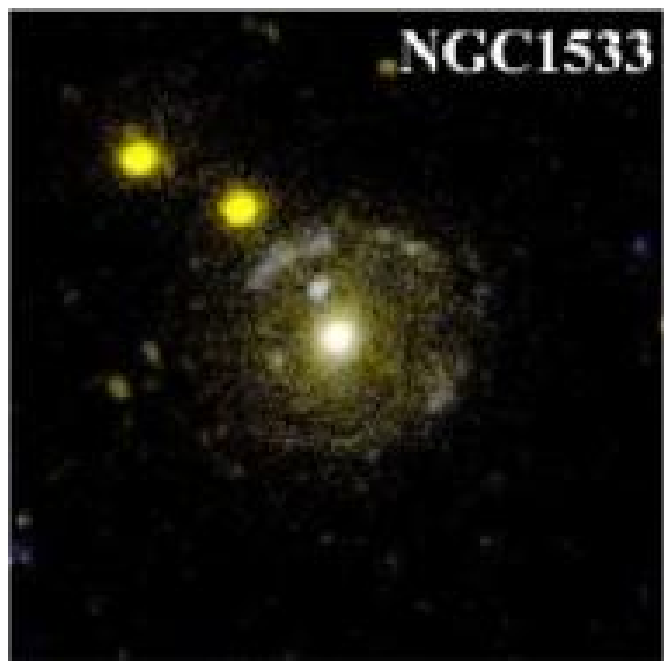}} &
{\includegraphics[width=4.5cm]{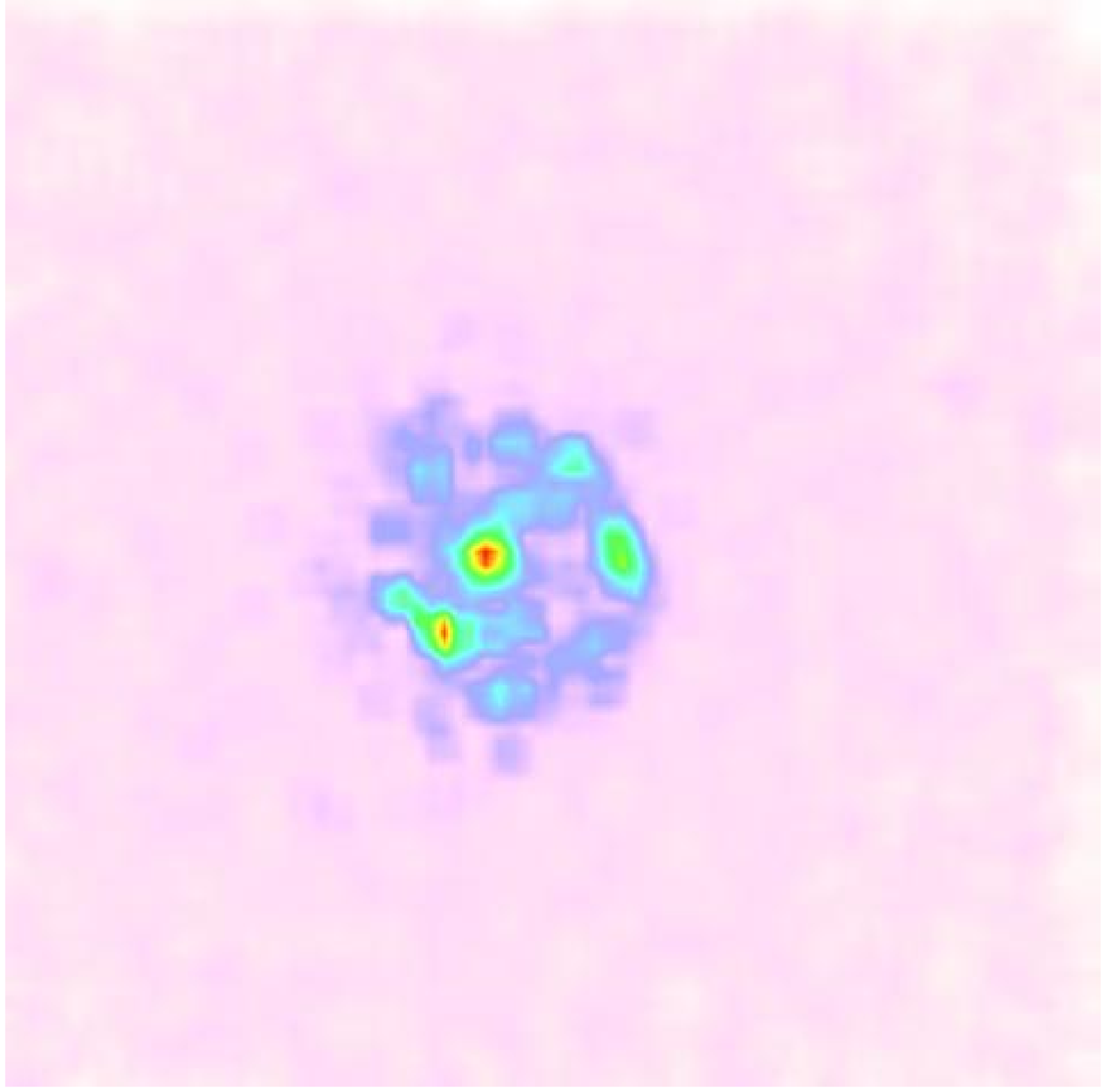}}\\
& {\includegraphics[width=4.5cm]{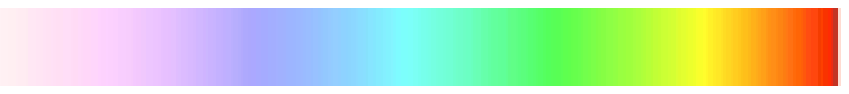}} \\
  {\includegraphics[width=5.5cm]{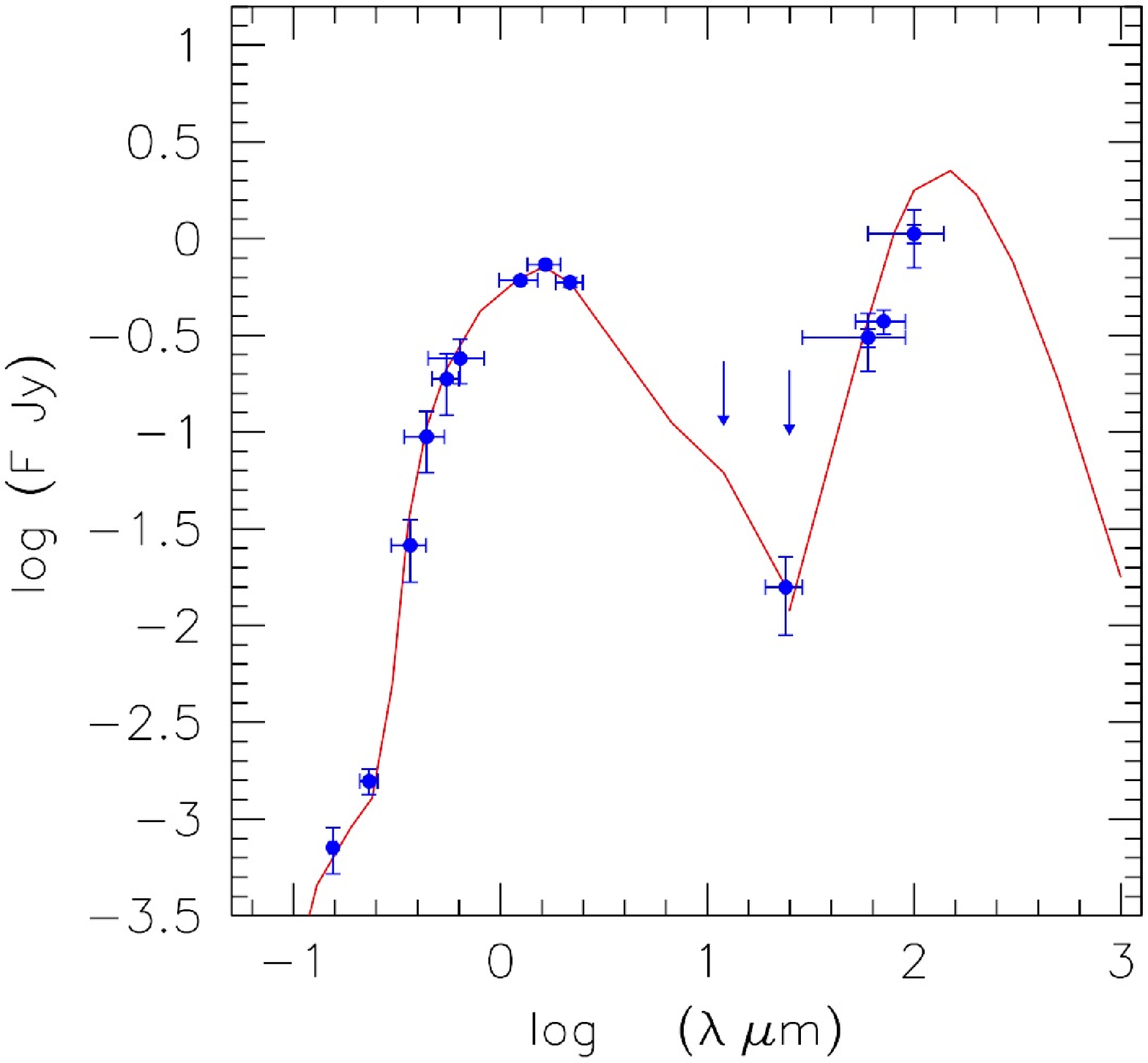}}&{\includegraphics[width=6cm]{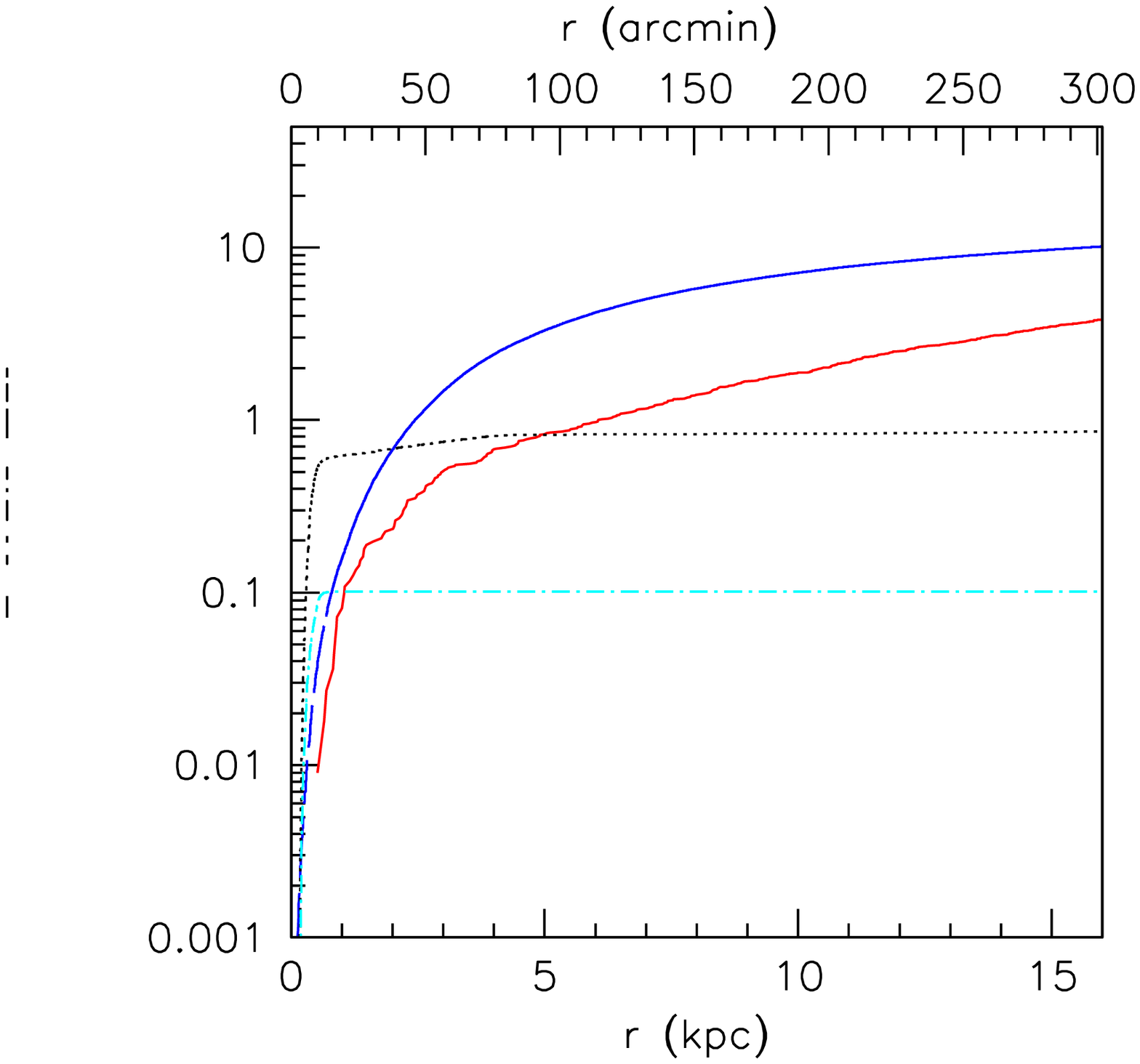}}\\
   \end{tabular}
 \caption{ {\sl Top:}  5'$ \times$5' colour composite UV (FUV blue, NUV yellow) image of  
 NGC~1533 (left) and  the simulated YZ projection FUV/NUV map (right),  on the same scale.  Flux has been normalized  to the total flux. 
The map includes  60 equally spaced levels  with a density contrast of 20. {\sl Bottom:} FUV to far-IR SED  of NGC~1533 (left);
blue points are the measured total fluxes  from {\it NED}; error bars account for band width and 3\,$\sigma$  uncertainties of the flux;  red  line  is the prediction of our model. {\sl Bottom right:} 
Mass distribution inside a radius of 16\,kpc $\simeq$10~$r_e$ centered on the V-band
galaxy centre (right); symbols are as in the left panel of Fig. \ref{fig20}.
}
      \label{1bis}
   \end{figure*}

Figure \ref{1bis} (top panels) compares, on the same scale, the observed composite 
FUV and NUV   image (left) with the selected snapshot (right). 
Figure \ref{1bis} (bottom left panel)  presents both the  observed SED of NGC~1533 
and the best-fit obtained from the selected simulation. The SED refers to the total flux of the galaxy.
FUV and NUV total magnitudes are derived from Table 3 of \citet{Ant11} and 
corrected for foreground galactic extinction following prescriptions in {\tt NED} \citep{F99}.

The entire SED is well matched at the selected snapshot.
The predicted FIR SED is composed of warm and cold dusts  as discussed in Sect \S 4.1.1.
The  intensity of the diffuse radiation  field, I$_0$=7I$_{local}$, 
is the same as in the Milky Way \citep{paola2}. 
The warm dust temperature is well constrained by {\tt MIPS} data of \citet{Temi09I}  at 24 and 70\,$\mu$m.
Our fit provides a warm dust temperature of 52~K  and a warm-to-cold energy ratio 0.15. 
Therefore, also in this case, the SED in the FIR is accounted for by the diffuse radiation field 
of disk stars on dust grains since cold dust accounts for 85$\%$ of FIR emission. 
As for NGC~3626, a small bulge characterizes this S0; its bulge-to-total light ratio is 0.25  \citep{Laurietal06}.
 The fraction of bolometric luminosity which comes out in the FIR,  
10\%, is 3 times less than  that expected for Sps but five times more than the average for S0s \citep{paola6,paola4}.

The total mass of the galaxy inside D$_{25}$  is 4.9$\times$10$^{10}$\,M$_{\odot}$ with a 
DM fraction of 0.17, and a mass-to-light ratio  in the B-band of 25.8\,M$_{\odot}$/L$_{\odot}$.

%-----------------------------------  Figure 1 --------------------------------
% ---------------------------- end Figure 1 --------------------------------   

Figure~\ref{1bis} (bottom right panel) shows the distribution of the different mass 
components at the selected snapshot within  16\,kpc, i.e.,  $\simeq$10~$r_e$. 
The  mass of cold gas, 1.0$\times$10$^9$\,M$_{\odot}$, is in good agreement with the 
findings by \citet{RW03}, given in Sect. \S 2.

\section{Insights into the evolution of NGC~3626 and NGC~1533}

\subsection{The (NUV-r) vs. Mr CMD}

As discussed by \citet{Schaw07}, (NUV-$r$) colour is an excellent tracer of
even small amounts ($\simeq$1\% mass fraction) of recent ($\le$1 Gyr)
star formation. In this context, the (NUV-$r$) CMD has been used
to study the effect of environment on the recent star formation
history in large sets of galaxies observed with {\it GALEX} \citep{Salimetal05}.

\begin{figure}
{\includegraphics[width=10cm]{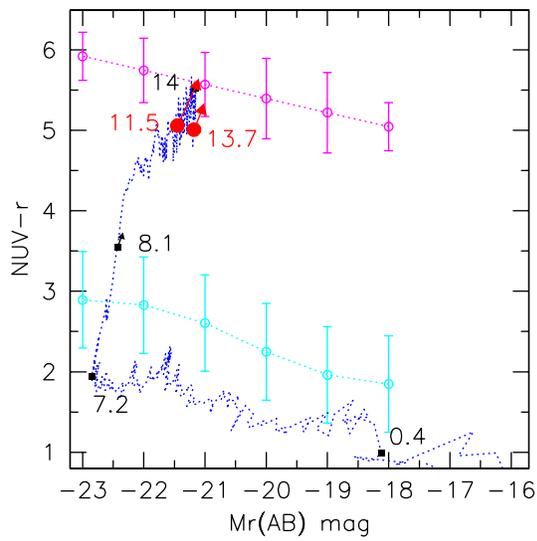}}
 \caption{The (NUV-r) vs M$_r$ CMD of our simulation best-fitting the 
 global properties of NGC~3626 and  NGC~1533. Dotted line  shows its evolutionary path; filled squares mark some meaningful 
 evolutionary times in Gyr;(red) filled circles show the current position of NGC~3626 and 
  NGC~1533;  arrows emphasize the  reddening effect. The \citet{Wyderetal07} fits of the 
 blue  (cyan) and red (magenta) sequences are plotted including their error bars.}
  \label{CMD}
\end{figure}

Our approach allows us to understand how the galaxy  transforms, i.e, what is its evolutionary   path in this diagram
(Fig. \ref{CMD}). 

\begin{figure*}
{\includegraphics[width=6.5cm]{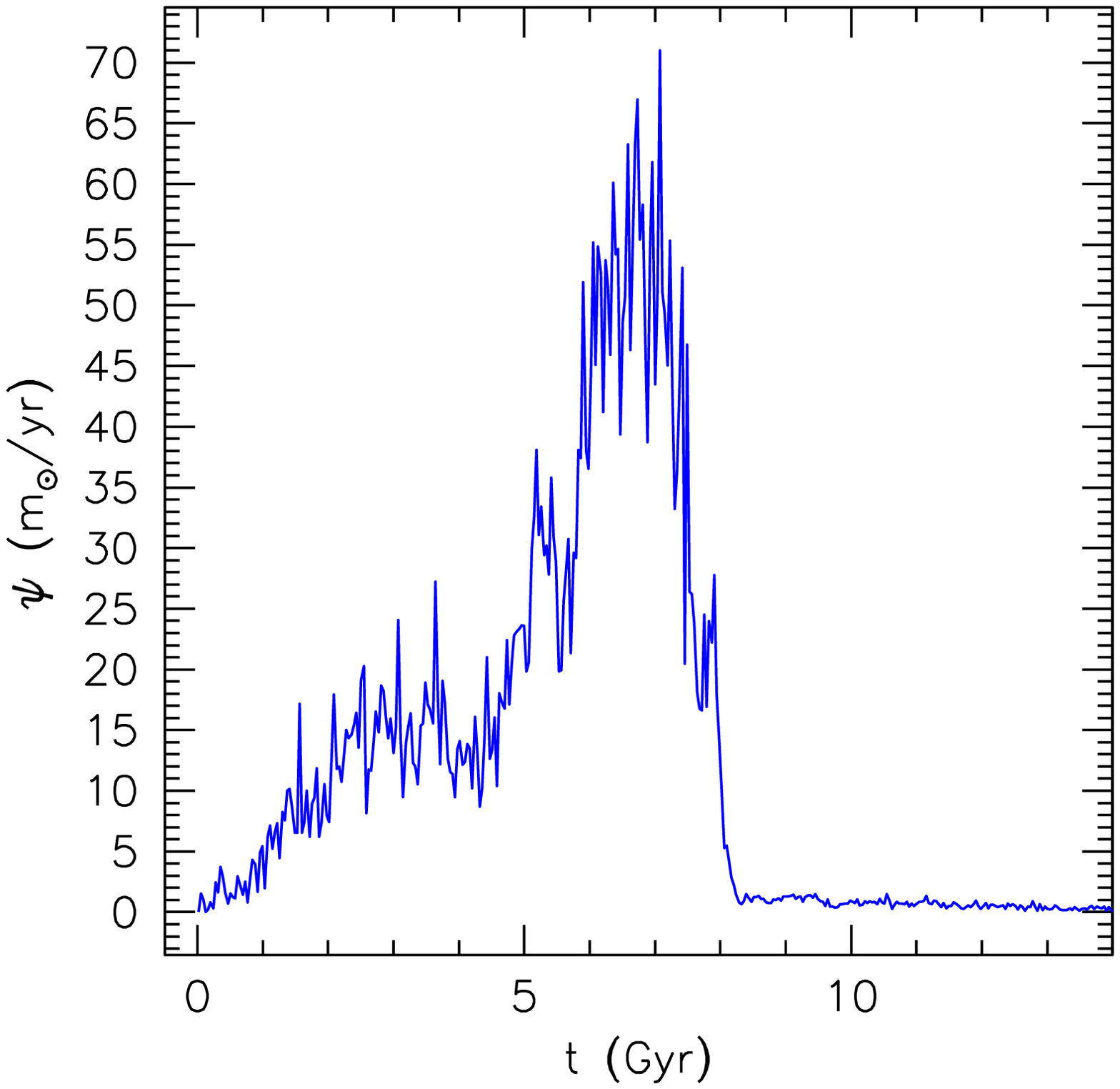}}
{\includegraphics[width=6.6cm]{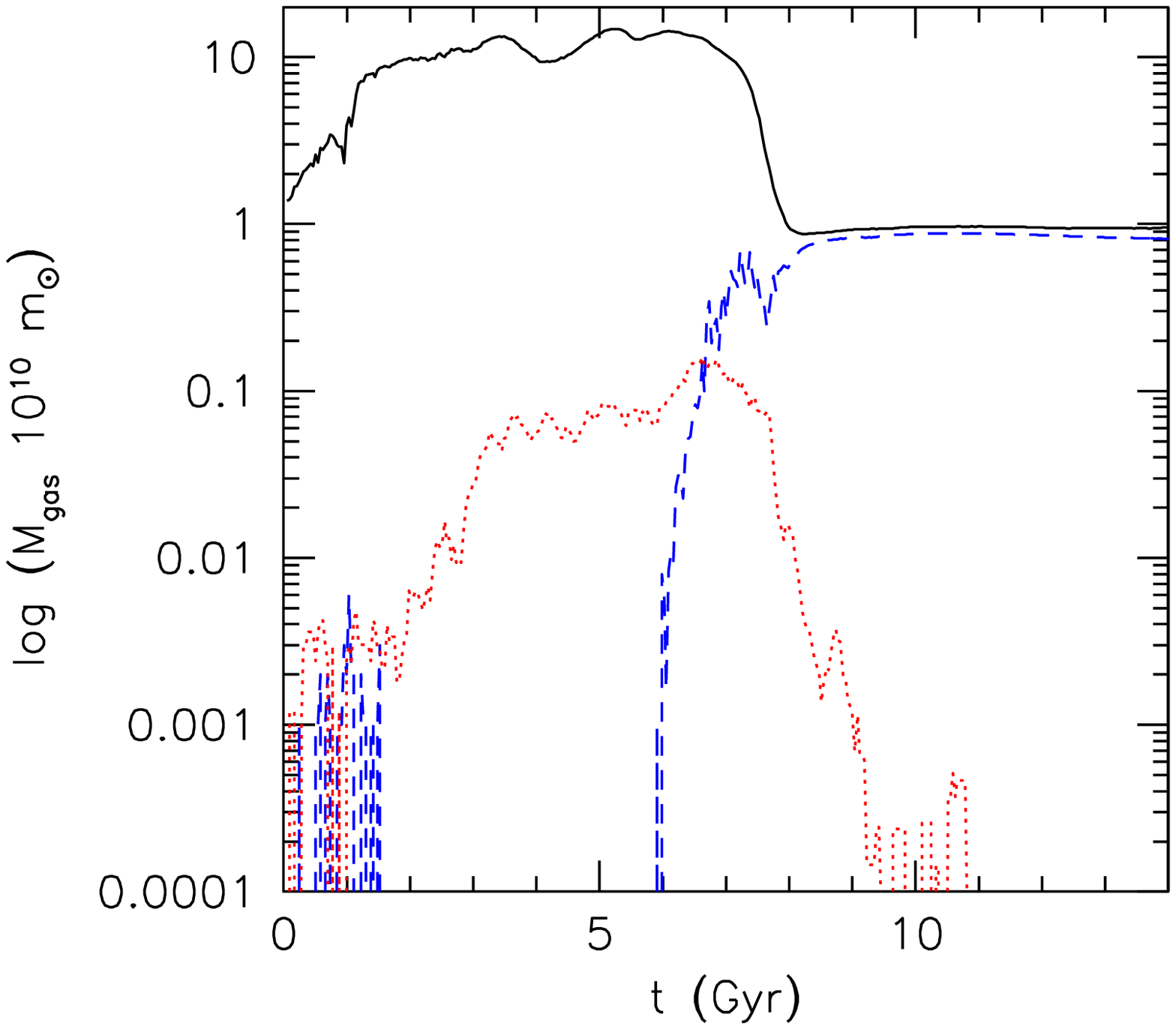}}
 \caption{ 
{\sl Left:} The time-evolution of the total SFR.
{\sl Right:}   (black) solid line  shows the evolution of the gas mass inside a 
 radius of 50\,kpc centered on the V-band luminous centre of the galaxy; 
 (blue) dashed line corresponds to the gas with temperature $\le$2$\times$10$^4$\,K, and the
 (red)  dotted line to the gas with temperature $\ge$  10$^6$\,K.
   }
 \label{fig3}
   \end{figure*}

The simulated galaxy gets to the blue  sequence \citep{Wyderetal07} after 0.4\,Gyr from the beginning, 
and follows it for 7.2\,Gyr, when reaching its maximum SFR (Fig. \ref{fig3}, left) and,
 correspondingly, its brighter M$_r$(AB) magnitude. 
At this point of the evolution, only 12\% of the galaxy bolometric luminosity  is absorbed and 
re-emitted in the FIR and  the galaxy is about 2 magnitudes  brighter in the B-band than at the 
 age of our best-fit for NGC 3626, i.e., 11.5\,Gyr.

Figure \ref{fig3} (left) shows that the total SFR  fades as the gas fueling decreases 
(right).  Moreover, its active phase lasts  7.2\,Gyr, then the SFR turns off  within 1\,Gyr. 
Correspondingly, as shown in Fig. \ref{CMD}, the galaxy leaves the blue sequence 
within 1\,Gyr. Then it crosses the GV to achieve the current position of NGC 3626 in  
3.4\,Gyr, and  stays in the red  sequence up to the end of our simulation ($\approx$14\,Gyr).

The effect of dust attenuation,  i.e. the internal reddening, evaluated accounting for an 
inclination angle of $\sim$60$^\circ$ (i.e. 56$^\circ$ for  NGC 3626, and 64$^\circ$ for 
NGC 1533 ({\tt HYPERLEDA}), does not modify this picture.
The attenuation, indeed, is quite negligible until the galaxy is 7\,Gyr old, due to the active role of 
the SFR (Fig. \ref{fig3}, left) which reduces the  cold gas (Fig. \ref{fig3}, 
 right). The internal reddening rises to A$_{(NUV-r)}$=0.13 mag
and A$_r$=0.07 mag at 8.1\,Gyr, it increases to 0.52 and 0.30 mag respectively at 11.5\,Gyr, 
the age of  NGC 3626, and reduces to 0.24 and 0.13 mag at 13.7\,Gyr, 
the age of  NGC 1533. The reddening is quite negligible at 14\,Gyr, the last 
point in Fig. \ref{CMD}. 
This finding agrees with the value of attenuation for ETGs \citep{Salimetal05}, i.e. their
attenuation decreases as their (NUV-$r$) colour becomes redder.
Moreover, the amount of reddening  we predict in the GV and red-sequence
regions of the CMD, agrees, within the errors, with the average value estimated by 
\citet[and references therein]{Martin07} even if \citet{Martin07} used the 
\citet{C94} law, whereas we adopt our own Galaxy extinction law
\citep{paola2}.

The simulation here does not predict strong oscillations in the GV, however 
different evolutionary paths in such a CMD are possible for early-type galaxies,
 as will be discussed in a forthcoming paper (Mazzei et al. 2013, in prep).

\subsection{Photometric evolution}
\begin{figure*}
  \centering
 {\includegraphics[width=6cm]{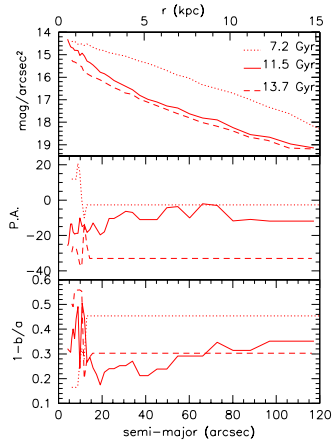}}
      \caption{{\sl Top:} compares the  J-band brightness profiles of our simulation at 7.2\,Gyr, when the  galaxy leaves the blue sequence
 (dotted line), at 11.5\,Gyr (solid line), i.e. the best-fit age of NGC~3626,  and at 13.7\,Gyr,
the best-fit age of NGC 1533. All the profiles are derived from the XZ projection of the corresponding snapshot with the same
resolution as in Fig. \ref{fig1} (top right) and Fig. \ref{fig20} (right).
{\sl Middle:} 
evolution of the position angle; symbols are the same
as in the top panel. {\sl Bottom:} evolution of the ellipticity.
}
 \label{fig6}
   \end{figure*}

Figure \ref{fig6} compares, at three meaningful evolutionary ages, the isophotal J-band
 brightness profiles (XZ projections) of the corresponding snapshots,  position angles 
 and ellipticities as derived from simulated maps with the same resolution as that 
  in Fig. \ref{fig1} (top right panel) and Fig. \ref{fig20} (right panel). 
Taking into account that our approach aims to reproduce the global properties 
of the galaxy, not all the details, Figure \ref{fig6} shows that a disk profile
 characterizes the  galaxy  morphology at ~7\,Gyr, when the galaxy is located on the 
 blue sequence of the CMD (Fig. \ref{CMD}). 
Therefore,  accounting for results in Sect. \S 4.1.1, a bulge progressively appears, and
  an outer disk is maintained during the evolution, i.e., along the GV up to the red sequence.  
  The position angle of the isophotes changes with  time, although it does not show any 
  significant twisting. 

% ------------------------ Figure 6 ---------------------------------
%----------------------- end Figure 6 ---------------------------------- 

Table \ref{tab:res1} summarizes galaxy predictions inside  a region of 11 
kpc, corresponding to the  R$_{25}$ radius of NGC~3626,  as a function of the ages 
reported in column~1.
We point out that  the r$_e$ of NGC~3626, 
as well as its R$_{25}$, are about twice those of NGC~1533 (Table \ref{tab:obs_e}).
The galaxy properties reported in Table \ref{tab:res1} are:  the average 
age of the stellar populations weighted by the B-band luminosity,  t$_{gen}$ (column 2),  the total mass 
(column 3), the mass-to-light ratio (column 4), the gas and DM mass fractions (column 5 and
6 respectively).

Table \ref{tab:res1} provides the following picture.  Inside the radius considered, 
the DM  is a low, constant fraction of the total mass, corresponding to less than 20\%, 
during all the evolution. The gas fraction decreases at the  age of the best-fit of 
NGC 3626, then remains constant. It shows that the gas accretion rate and the SFR 
balance themselves within the radius considered. The mass-to-light-ratio increases 
with  age following the decrease of  the galaxy luminosity in the B-band.
 
%-------------------------- Table 4 ------------------------------------------
\begin{table*}
\caption{Results within 11\,kpc}
\begin{tabular}{lccccccc}
\hline
galaxy age & t$_{gen}$  & M$_{tot}$& M/L$_B$ & f$_{gas}$ & f$_{DM}$\\ 
Gyr & Gyr& 10$^{10}$\,M$\odot$ & M$\odot$/L$\odot$& &\\
\hline
\hline
7.2    & 0.6 & 11.0 &2.3& 0.22 & 0.19  \\
11.5   & 4.5 & 11.5 &18.8& 0.08 & 0.19  \\
13.7  & 7.0 & 10.5 & 33.2& 0.08 & 0.19 \\
\hline
\end{tabular}

{\footnotesize {t$_{gen}$: the average age of stellar population weighted on B-band luminosity}}

\label{tab:res1}
\end{table*}
%-------------------------- end Table 4 -------------------------------------

\section{Discussion and Conclusions}

We are exploring the co-evolution of  galaxies in groups combining 
multi wavelength photometric and 2D kinematical observations \citep{Ant12}. 
S0s in groups are of particular interest since  the possible transformation 
Sp$\rightarrow$S0 may be tested in an environment physically
different from that of a cluster. We simulate the evolution of Sa/S0 galaxies
showing  SF in   ring/arm like structures both detected by 
H$\alpha$ and/or FUV observations and characterized by the presence 
of large-scale  HI structures  in and/or around them. 

 These ETGs are not so uncommon in low 
density environments, like groups. \citet{Serra10} found that $\geq25$\% of ETGs 
contain HI at the level of M(HI)$\geq 10^8$\, M$_\odot$, i.e. that M(HI) is of the order of a few per cent of the total stellar mass. 
Their HI interferometric observations revealed very disturbed gas 
morphology/kinematics in a vast majority of the detected systems, confirming the 
continuing assembly of many ETGs but also showing that these are not necessarily gas free. 
They also find that all very disturbed ETGs have a single-stellar-population-equivalent age 
$<4$ Gyr.

Here we focus on two of such galaxies, NGC~3626 and NGC~1533. 
Our SPH simulations allow us to derive dynamical and morphological information 
as well as the SED extending over four orders of magnitude at each evolutionary stage.

We find that the global properties of both NGC~1533 and NGC~3626 are well matched
 by the same simulation, suggesting that their complex properties may be explained in 
 the framework of 
galaxy encounters. In the specific case the driver of both the photometric and kinematics 
peculiarities of NGC 3626 and NGC 1533 is a major merging episode. 
The merger occurs from haloes initially composed of DM and gas with  mass ratio 2:1,
total mass 3$\times$10$^{12}$\,M$_{\odot}$,  and gas fraction 0.1.
Star forming ring/arm like structures observed in the galaxy disk are  features 
arising in the latter stages of the merger episode  which starts after 3.5\,Gyr from the onset of the SF.
These features arise when the galaxy is almost 8\,Gyr old,
following the head-on collision driving the evolution of our systems \citep{MB09}.
NGC~3626 is 11.5\,Gyr old, and NGC~1533 13.7\,Gyr old. 
Their FIR SED is dominated by the emission of cold dust, heated by the diffuse radiation
 field which represents, in both the cases, at least 80\% of the total FIR emission. 
 Moreover, the fraction of bolometric luminosity which comes out in the FIR spectral
range is  five times  and ten times  larger for NGC~1533 and NGC~3626 respectively, 
than the average for S0 galaxies \citep{paola6}. Using new 
{\tt AKARI/FIS} catalogue data \citep{Yamaetal09}, we conclude that NGC 3626 is characterized 
by almost  the same FIR fraction as the Milky Way, i.e. 30\%  \citep{paola2}.

Our  chemo-photometric SPH simulations, including
evolutionary population synthesis models, allow us to  trace,  in a fully consistent way, the evolutionary path of  
NGC 3626 and NGC 1533 in the  {\it (NUV-r) - $M_r$} CMD. 
The transition from the blue sequence, along which these galaxies 
evolved for about  7-8~Gyr, across the GV and before they became mature and
red ETGs, lasts about 4-5 Gyr. \citet{Fasano2000} noticed that a sort of  morphological 
  conversion  in the cluster galaxy population from Sp$\rightarrow$S0 took place 
about 1 - 4 Gyr ago. This  estimate agrees with the time scales of our simulations. 
Excursions into the GV from the red sequence, driven by the acquisition of fresh
gas for star formation \citep{Thilker10}, are not expected in the evolutionary picture 
characterizing these two galaxies, but are possible  along different evolutionary 
paths (Mazzei et al. 2013, in prep). 
The global properties found  for NGC~1533 and NGC~3626  arise from  
the same evolutionary framework already explored in \citet{Dani12}: a merging episode with a  gas rich companion.

 Several mechanisms have been proposed to explore transformation of 
Sp$\rightarrow$S0 in groups, from {\it strangulation} \citep{Kawata08}, to repetitive slow
encounters \citep{Bekki11} both grabbing gas from  Sp, quenching star formation
and suppressing arms.  Our results show that another mechanism of gravitational origin, the
merging of two haloes, is a {\it viable} mechanism to generate  S0s today found 
in the red sequence, in agreement with suggestions by \citet {Boselli06} and \citet{Getal13}.

\section*{Acknowledgments}
We thank Martha Haynes for kindly sending us the  H$\alpha$ image of NGC 3626 and Marcel Clemens for help in revising the text. 
RR, AM and DB acknowledge the partial financial support by contract ASI-INAF  I/009/10/0.
DB and RR acknowledge the partial financial support by contract INAF/PRIN 2011 
``Galaxy Evolution with the VLT Survey Telescope (VST)''.
This research has made use of the NASA/IPAC Extragalactic Database (NED) which is operated 
by the Jet Propulsion Laboratory, California Institute of Technology, under contract with the 
National Aeronautics and Space Administration. 
We acknowledge the usage of the HyperLeda database (http://leda.univ-lyon1.fr).

%%%%%%%%%%%%%%%%%%%%%%%%%%%%%%%%%%%%%%%%%%%%%%%%%%%%%%%%%%%%%%%%%%%%%%%%%%%%%


\begin{thebibliography}{}

\bibitem[\protect\citeauthoryear{Afanesiev \& Sil\'chenko}{2007}]{AS07}Afanesiev, V., Silchenko ,O., 
{\it Leo II Group: decoupled cores of NGC 3607 and NGC 3608}, A\&AT, 26, 311-337, 2007
\bibitem[\protect\citeauthoryear{Annibali et al.}{2007}]{Ann07}Annibali, F., Bressan, A., Rampazzo, R. et al., {\it Nearby early-type galaxies with ionized gas III. Analysis of line-strength indices with new stellar population models}, A\&A, 463, 455-479, 2010
\bibitem[\protect\citeauthoryear{Baldry et al.}{2004}]{Baldry04}Baldry I.~K., Glazebrook, K., Brinkmann, J., Ivezi{\'c}, {\v Z}., Lupton,  R.~H., Nichol, R.~C., \& Szalay, A.~S. , {\it Quantifying the Bimodal Color-Magnitude Distribution of Galaxies}, ApJ, 600, 681-694, \ 2004
\bibitem[\protect\citeauthoryear{Bekki}{2009}]{Bekki09} Bekki, K. {\it Ram-pressure stripping of halo gas in disc galaxies: implications for galactic star formation in different environments}, MNRAS, 339, 2221-2230, 2009  
\bibitem[\protect\citeauthoryear{Bell}{2003}]{Bell03}Bell, E.F., {\it Estimating star formation rates from infrared and radio luminosities: the origin of the radio-infrared correlation}, ApJ, 586, 794-813, 2003.
\bibitem[\protect\citeauthoryear{Bekki \& Couch}{2011}]{Bekki11}Bekki, K, Couch, W. 
{\it Transformation from spirals into S0s with bulge growth in groups galaxies}, MNRAS, 415, 
1783-1796, 2011
\bibitem[\protect\citeauthoryear{Bettoni et al.}{2012}]{Dani12}Bettoni, D., Buson, L., Mazzei, P., Galletta, G., {\it Insight into the evolution of the innermost region of the NGC 1023 Group}, MNRAS, 423, 2957-2965,\ 2012
\bibitem[\protect\citeauthoryear{Binggeli \& Hascher}{2007}]{BH07}Binggeli, B.\& Hascher, T., {\it Is There a Universal Mass Function?}, PASP, 119, 592-604, \ 2007 
\bibitem[\protect\citeauthoryear{Boselli \& Gavazzi}{2006}]{Boselli06} Boselli, A, Gavazzi G. 
{\it Environmental effects on late-type galaxies in nearby clusters}, PASP, 118, 517-559, 2006
\bibitem[\protect\citeauthoryear{Calzetti et al.}{1994}]{C94}Calzetti, D., Kinney, A. \& Storchi-Bergmann, T.\ {\it Dust extinction of the stellar continua in starburst galaxies: The ultraviolet and optical extinction law},  ApJ, 429, 582-601, 1994
\bibitem[\protect\citeauthoryear{Ciri et al.}{1995}]{Cetal95}Ciri, R., Bettoni, D., Galletta, G., {\it A massive counter-rotating gas disk in a spiral galaxy}, Nature, 375, 661-663, \ 1995
\bibitem[\protect\citeauthoryear{Condon et al.}{2002}]{cond02}Condon, J., Cotton, W., Broderick, J., {\it Radio Sources and Star Formation in the Local Universe}, AJ, 164, 675-689, \  2002
\bibitem[\protect\citeauthoryear{Cortese \& Hughes}{2009}]{Cortese09}{\it Evolutionary path 
to and from the red sequence: star formation and HI properties of transition galaxies at z$\sim$0},
MNRAS, 400, 1225-1240, 2009
 \bibitem[\protect\citeauthoryear{Cram et al.}{1998}]{Crametal}Cram,L., Hopkins, A., Mobasher, B., Rowan-Robinson, M., {\it Star Formation Rates in Faint Radio Galaxies}, ApJ, 507, 155-160,\ 1998
\bibitem[\protect\citeauthoryear{Curir \& Mazzei}{1999}]{CM99}Curir, A.,  Mazzei, P. {\it SPH simulations of galaxy evolution including chemo-photometric predictions},  New Astron., 4, 1-20, \ 1999
\bibitem[\protect\citeauthoryear{Dekel \& Birnboim}{2006}]{Dekel06}Dekel, A., Birnboim, Y.
{\it Gas bimodality due to cold flows and shock heating}, MNRAS, 368, 2-20, 2006
\bibitem[\protect\citeauthoryear{de Vaucouleurs et al.}{1991}]{DV91} de Vaucouleurs, G., de Vaucouleurs, A., Corwin, H. G., Buta, R. J., Paturel,
G.,  Fouque, P., {\it Third Reference Catalogue of Bright Galaxies}, 1991 (New York : Springer) (RC3)
\bibitem[\protect\citeauthoryear{DeGraaff et al.}{2007}]{dG07}DeGraaff, R., Blakeslee, J., Meurer, J. {\it A galaxy in transition: structure, globular clusters, and distance of the star-forming S0 galaxy NGC 1533 in Dorado}, ApJ, 671, 1624-1639, 2007
\bibitem[\protect\citeauthoryear{Fang et al.}{2012}]{Fang12} Fang, J.J., Faber, S.M., Salim, S., Graves, G.J., Rich, R.M., {\it The slow death (or rebirth?) of extended star formation in z=0.1 green valley early-type galaxies}, ApJ, 676, 23-39, 2012
\bibitem[\protect\citeauthoryear{Fasano et al.}{2000}]{Fasano2000} Fasano, G., Poggianti, B., Couch, W., Bettoni, D., Kjaergaard, P., Moles, M., {\it The Evolution of the Galactic Morphological Types in Clusters}, ApJ, 542, 673-683,  2000
\bibitem[\protect\citeauthoryear{Freeman}{1970}]{F70}Freeman, K.\,C., {\it On the disks of Spiral and S0 galaxies}, ApJ, 160, 811-830, 1970
\bibitem[\protect\citeauthoryear{Fitzpatrick}{1999}]{F99}Fitzpatrick, E., {\it Correcting for the Effects of Interstellar Extinction}, PASP, 111, 63-75, 1999.
\bibitem[\protect\citeauthoryear{Garc\'ia-Burillo et al.}{1998}]{G98} Garc\'ia-Burillo, S., Sempere, M., Bettoni, D.\ {\it First Detection of a Counter rotating Molecular Gas Disk in a Spiral Galaxy: NGC 3626}, ApJ, 502, 235-244, 1998
\bibitem[\protect\citeauthoryear{George et al.}{2013}]{Getal13}George, M., Chung-Pei, M., Bundy, K. et al. {\it Galaxies in X-ray Groups. III. Satellite Color and Morphology Transformations}, arXiv:astro-ph/1302.6620, 1-11, 2012
\bibitem[\protect\citeauthoryear{Gil de Paz et al.}{2007}]{Gil07}Gil de Paz, A., Boissier, S., Madore, B., et al. {\it The GALEX Ultraviolet Atlas of Nearby Galaxies}, ApJS, 173, 185-255, 2007
\bibitem[\protect\citeauthoryear{Grossi et al.}{2009}]{Grossi09} Grossi, M., di Serego Alighieri, S., Giovanardi, C. et al., {\it The HI content of early-type galaxies from the ALFALFA survey. II. The case of low density environments}, A\&A 498, 407-417, 2009
\bibitem[\protect\citeauthoryear{Haynes et al.}{2000}]{Haynesetal00}Haynes, M., Jore. K., Barrett, E.,  et  al. \ {\it Kinematic Evidence of Minor Mergers in Normal SA Galaxies: NGC 3626, NGC 3900, NGC 4772, and NGC 5854}, ApJ, 120, 703-727, 2000.
\bibitem[\protect\citeauthoryear{Hopkins et al.}{2003}]{Hop03} Hopkins, A., Miller, C., Nichol, A., et al. {\it Star Formation Rate Indicators in the Sloan Digital Sky Survey}, ApJ, 599, 971-991, 2003
\bibitem[\protect\citeauthoryear{Kawata \& Mulchaey}{2008}]{Kawata08} Kawata, D., Mulchaey, J.S, {\it Strangulation in galaxy groups}, ApJ, 672, L103-L106, 2008
\bibitem[\protect\citeauthoryear{Horellou et al}{2001}]{Horellou01} Horellou, C., Black, J. H., van Gorkom, J. H., Combes, F., van der Hulst, J. M., Charmandaris, V., {\it Atomic and molecular gas in the merger galaxy NGC 1316 (Fornax A) and its environment}, A\&A, 376, 837-852, 2001
\bibitem[\protect\citeauthoryear{Hughes \& Cortese}{2009}]{Hughes09} Hughes, T.M., Cortese, L.
{\it The migration of nearby spirals from the blue to the red sequence: AGN feedback or environmental effects?}, MNRAS, 396, L41-L45, 2009
\bibitem[\protect\citeauthoryear{Kroupa}{2012}]{K12} Kroupa, P. {\it The Dark Matter Crisis: Falsification of the Current Standard Model of Cosmology}, PASA, 29, 395-433, 2012
\bibitem[\protect\citeauthoryear{Jedrzejewski}{1987}]{Jedr87} Jedrzejewski R. I., {\it CCD surface photometry of elliptical galaxies. I - Observations, reduction and results}, MNRAS, 226, 747-768, 1987
\bibitem[\protect\citeauthoryear{Just et al.}{2010}]{Just10}Just, D. W., Zaritsky, D., Sand, D. J., Desai, V.,  Rudnick, G., {\it The Environmental Dependence of the Evolving S0 Fraction}, ApJ, 711, 192-200, 2010
\bibitem[\protect\citeauthoryear{Laurikainen et al.}{2005}]{Laurietal05} Laurikainen, E., Salo, H., \& Buta, R., {\it Multicomponent decompositions for a sample of S0 galaxies},  MNRAS, 362, 1319-1347, 2005.
\bibitem[\protect\citeauthoryear{Laurikainen et al.}{2006}]{Laurietal06} Laurikainen, E., Salo, H., \& Buta, R., {\it Morphology of 15 southern early-type disk galaxies}, ApJ, 132, 2634-2652, 2006.
\bibitem[\protect\citeauthoryear{Lewis et al}{2002}]{Lewis02}Lewis, I., Balogh, M., De Propris, R., Couch, W., et~al., {\it The 2dF Galaxy Redshift Survey: the environmental dependence of galaxy star formation rates near clusters}, MNRAS, 334, 673-683, \ 2002
\bibitem[\protect\citeauthoryear{Martin et al.}{2007}]{Martin07}Martin, C., Wyder, T., Schiminovich, D., et~al., {\it The UV-optical galaxy color-magnitude diagram. III. Constraints on evolution from the blue to the red sequence}, ApJS, 173, 342-356, 2007.
\bibitem[\protect\citeauthoryear{Marino et al.}{2013}]{Ant12}Marino, A., Rampazzo, R., Bianchi, L., et al., {\it Galaxy evolution in nearby loose groups. II. Photometric and kinematic characterization of USGC U268 and USGC U376 group members in the Leo cloud},  MNRAS, 428, 476-501, 2013
\bibitem[\protect\citeauthoryear{Marino et al.}{2011a}]{Ant11_II}Marino, A., Bianchi, L., Rampazzo, R., et al., {\it Tracing rejuvenation events in nearby S0 galaxies}, ApJ, 736, 154-162, 2011a 
\bibitem[\protect\citeauthoryear{Marino et al.}{2011b}]{Ant11}Marino, A., Rampazzo, R., Bianchi, L., et al., {\it Nearby early-type galaxies with ionized gas: the UV emission from GALEX observations}, MNRAS, 411, 311-331, 2011b 
\bibitem[\protect\citeauthoryear{Mazzei et al.}{1992}]{paola2}Mazzei, P., Xu, C., \& de Zotti, G., {\it A model for the photometric evolution of disc galaxies from UV to far-IR}, A\&A, 256, 45-55, 1992.
\bibitem[\protect\citeauthoryear{Mazzei et al.}{1994}]{paola4}Mazzei, P. \&  De Zotti, G., Xu, C.,
{\it Models for the evolution of the spectral energy distribution of ellipticals galaxies from ultraviolet to far-IR wavelengths}, ApJ, 422, 81-91, 1994. 
\bibitem[\protect\citeauthoryear{Mazzei \& De Zotti}{1994}]{paola6}Mazzei, P. \& De Zotti, G., {\it The far-IR properties of early-type galaxies}, ApJ, 426, 97-104, 1994.
\bibitem[\protect\citeauthoryear{Mazzei}{2003}]{paola1}Mazzei, P.\, {\it The history of star formation in galaxies. Insights from SPH simulations of triaxial collapsing systems}, Mem. Sait, 74, 498-499, 2003.
\bibitem[\protect\citeauthoryear{Mazzei \& Curir}{2003}]{paola5}Mazzei, P. \&  Curir,  A., {\it Dark and luminous matter connections from smooth particle hydrodynamics simulations of isolated collapsing triaxial systems}, ApJ, 591, 784-790, 2003: MC03
\bibitem[\protect\citeauthoryear{Mazzei}{2004}]{paolaa}Mazzei, P.,{\it Dark and luminous matter connections. Toward understanding galaxy evolution}, Research Signpost, Recent Res. Devel. Astron. \& Astrop., 1, 457-473,  2003, (arXiv:astro-ph/0401509)
\bibitem[\protect\citeauthoryear{Moiseev \& Bizyaev}{2009}]{MB09}Moisee, A., Bizyaev, D., {\it 3D spectroscopic study of galactic rings: Formation and kinematics}, New Astron. Rev., 53, 169-174, 2009
\bibitem[\protect\citeauthoryear{Navarro et al.}{1997}]{NFW97}Navarro, J., Frenk, C, White, S., {\it A Universal Density Profile from Hierarchical Clustering},
ApJ, 490, 493-508, 1997
\bibitem[\protect\citeauthoryear{Paturel et al.}{2003}]{Patetal03} Paturel G., Petit C., Prugniel P., et al. {\it HYPERLEDA. I. Identification and designation of galaxies}, A\&A, 412, 45-55, 2003
\bibitem[\protect\citeauthoryear{Phillipps \&  Disney}{1983}]{PD83} Phillipps, S., Disney, M.,
{\it The surface brightness of spiral galaxies. I - Spheroidal components and Freeman's law}, MNRAS, 203, 55-65, 1983
\bibitem[\protect\citeauthoryear{Ramella et al.}{2002}]{Rametal02}Ramella, M., Geller, M., Pisani, A., da Costa L., 
{\it The UZC-SSRS2 Group Catalog}, AJ, 123, 2976-2984, 2002

\bibitem[\protect\citeauthoryear{Rampazzo et al.}{2013}]{Rampazzo13}Rampazzo, R., Panuzzo, P., Vega, O. et al. {\it A Spitzer-IRS spectroscopic atlas of early-type galaxies in the Revised Shapely-Ames Catalog}
 MNRAS, in press. (DOI 10.1093/mnras/stt475)
\bibitem[\protect\citeauthoryear{Rifatto et al.}{1995}]{Rifatto95} Rifatto, A., Longo, G., Capaccioli, M., {\it The UV properties of normal galaxies. III. Standard luminosity profiles and total magnitudes},
A\&A Sup. Ser., 114, 527-536, 1995
\bibitem[\protect\citeauthoryear{Ryan-Weber et al.}{2003}]{RW03}Ryan-Weber, E., Webster, R, Staveley-Smith, L. {\it The 1000 Brightest HIPASS galaxies: The HI mass function and $\Omega_{HI}$}, MNRAS, 343, 1195-1206, 2003
\bibitem[\protect\citeauthoryear{Salpeter}{1955}]{Salp55}Salpeter, E.E., {\it The Luminosity Function and Stellar Evolution.}, ApJ, 121, 161-167,\ 1955
\bibitem[\protect\citeauthoryear{Salim et al.}{2005}]{Salimetal05}  Salim, S., Charlot, S., Rich R.M., et al. {\it New constraints on the star formation History and dust attenuation of galaxies in the local Universe from {\it GALEX}} ApJ, 619, L39-L42, 2005
\bibitem[\protect\citeauthoryear{Salim et al.}{2012}]{Salim12}  Salim, S., Fang, J. J., Rich, R.M. et al. {\it Galaxy-scale star formation on the red sequence: the continued growth of S0s and the quiescence of ellipticals} ApJ, 755, 105(pp.29), 2012
\bibitem[\protect\citeauthoryear{Sandage \& Tammann}{1987}]{Sandage87} Sandage, A., Tammann, G. A. 
{\it A revised Shapley-Ames Catalog of bright galaxies}, Carnegie Institution of  Washington Publication, Washington: Carnegie Institution 2nd edition, 1987  (RSA)
\bibitem[\protect\citeauthoryear{Schawinski et al.}{2007}]{Schaw07}Schawinski, K., Kaviraj, S., Khochfar, S. et al., {\it The Effect of Environment on the Ultraviolet Color-Magnitude Relation of Early-Type Galaxies}, ApJS, 173, 512-523, \ 2007
\bibitem[\protect\citeauthoryear{Schawinski et al.}{2009}]{Schawinski09}Schawinski, K., Virani, S., 
Simmons, B., Urry, M.C., Treister, E., Kaviraj S., Kushkuley, B. {\it Do moderate-luminosity active galactic nuclei suppress star formation?} ApJ, 692, L19-L23, 2009 
\bibitem[\protect\citeauthoryear{Serra \& Oosterloo}{2010}]{Serra10} Serra, P., Oosterloo, T.A.
{\it Cold gas and young stars in tidally disturbed ellipticals at z = 0}, MNRAS, 401, L29–L33, 2010
\bibitem[\protect\citeauthoryear{Sil'chenko et al.}{2010}]{Sietal10}Sil\'chenko ,O., Moiseev, A., \& Shulga, A. , {\it Lenticular galaxies at the outskirts of the Leo II group: NGC 3599 and NGC 3626}, AJ,  140, 1462-1474, 2010.
\bibitem[\protect\citeauthoryear{Spavone et al.}{2009}]{marilena}Spavone, M., Iodice, E., Calvi, R., Bettoni, D., Galletta, G., Longo, G., Mazzei, P. \& Minervini, G. {\it Revisiting the formation history of the minor-axis dust lane galaxy NGC 1947}, MNRAS, 393, 317-328, 2009 
\bibitem[\protect\citeauthoryear{Spavone et al.}{2012}]{marilena2}Spavone, M., Iodice, E., Bettoni, D., Galletta,  Mazzei, P. \& Reshetnikov, V., {\it A new photometric investigation of the double-ringed galaxy ESO 474-G26: unveiling the formation scenario}, MNRAS, 426, 2003-2018,  2012
\bibitem[\protect\citeauthoryear{Strateva et al.} {2001}]{Strateva01}Strateva, I., Ivezi{\'c}, Knapp, et~al. {\it Color separation of galaxy types in the Sloan Sky Survey imaging data}, AJ, 122, 1861-1874, 2001
\bibitem[\protect\citeauthoryear{Temi et al.}{2009a}]{Temi09I}Temi, P., Brighenti, F., Mathews, W., {\it  SPITZER observations of passive and star-forming early-type galaxies: an infrared color-color sequence}, ApJ, 707, 890-902, 2009
\bibitem[\protect\citeauthoryear{Thilker et al.}{2007}]{Thilker07} Thilker, D. A., Bianchi, L.,  Meurer, G. et al. 
{\it A search for extended ultraviolet disk (XUV-disk) galaxies in the local Universe}, ApJS, 714, 538-571, 2007
\bibitem[\protect\citeauthoryear{Thilker et al.}{2010}]{Thilker10} Thilker, D. A., Bianchi, L.,  Schiminovich, D. et al. 
{\it NGC 404: A Rejuvenated Lenticular Galaxy on a Merger-induced, Blueward Excursion Into the Green Valley}, ApJ, 714, L171-L175, 2010
\bibitem[\protect\citeauthoryear{Villalobos et al.}{2012}]{Villalobos12}
Villalobos, A., De Lucia, G., Borgani, S, Murante, G. {\it Simulating the evolution of disc galaxies in group environment.I. The influence of the global tidal field}, MNRAS, 424, 2401-2428, 2012
\bibitem[\protect\citeauthoryear{Warren et al.}{1992}]{war92}Warren, M., Quinn, P., Salmon, J. \& Zurek, W. \ {\it Dark halos formed via dissipationless collapse. I - Shapes and alignment of angular momentum}, ApJ, 399, 405-425, 1992
\bibitem[\protect\citeauthoryear{Weijmans et al.}{2008}]{Weijmans08} Weijmans, A-M., Krajnovi\`c,  D., van de Ven, G. et al., {\it The shape of the dark matter halo in the early-type galaxy NGC 2974}, MNRAS, 383, 1343-1358, 2008
\bibitem[\protect\citeauthoryear{Werk et al.}{2010}]{Werketal10} Werk, J, Putman, M, Meurer, G. et al. {\it Outlying HII regions in HI-selected galaxies}, AJ, 139, 279-295, 2010. 
\bibitem[\protect\citeauthoryear{Wetzel et al.}{2012}]{Wetzeletal12}Wetzel, A., Tinker, J., Conroy, C. et al. {\it Galaxy evolution in groups and clusters: satellite star
formation histories and quenching timescales in a hierarchical Universe}, MNRAS, 1-24, 2012
\bibitem[\protect\citeauthoryear{Wilman et al.}{2009}]{Wilman09}Wilman, D. J., Oemler, A., Mulchaey, J. S., McGee, S. L., Balogh, M. L., Bower, R. G., {\it Morphological Composition of z ~ 0.4 Groups: The Site of S0 Formation}, ApJ, 692, 298-308, 2009
\bibitem[\protect\citeauthoryear{Wyder et al.}{2007}]{Wyderetal07}Wyder, T., Martin, K., Schiminovich, D. et al,
 {\it The UV-Optical Galaxy Color-Magnitude Diagram. I. Basic Properties}, ApJSS, 173, 293-314,  2007
\bibitem[\protect\citeauthoryear{Yamamura et al.}{2009}]{Yamaetal09}Yamamura, I., Makiuti, S., Ikeda, N., et al.\ {\it The first release of the AKARI/FIS  Bright source catalogue}, AIP Conf. Proc. {\sl The exoplanets and disks: their formation and diversity}, 1158, 169-170, 2009

\end{thebibliography}
\end{document}